\documentclass[12pt,a4paper]{article}
\usepackage{jheppub}  
\usepackage{amssymb} 
\usepackage{amsmath}
\usepackage{mathtools}
\usepackage{amsfonts}    
\usepackage{dsfont}
\usepackage{pdfpages}
\usepackage{verbatim}
\usepackage{tensor}
\usepackage{tikz}
\usetikzlibrary{arrows,shapes,positioning,decorations.markings,calc}
\usepackage[tikz]{mdframed}
\tikzstyle arrowstyle=[scale=1]
\tikzstyle directed=[postaction={decorate,decoration={markings,
    mark=at position .55 with {\arrow[arrowstyle]{stealth}}}}]
\usepackage{ifthen}
\usepackage{multirow}
\usepackage{longtable}
\usepackage[vcentermath]{youngtab}
\usepackage{afterpage}
\usepackage{lscape}
\usepackage{rotating}
\usepackage{float}

\makeatletter
\def\@fpheader{\vspace{0cm}}
\makeatother

\newif\ifanswers
\answerstrue

\newcounter{question}
\stepcounter{question}
\newcommand{\questionnumber}[1]{\addtocounter{question}{-1}\refstepcounter{question}\label{#1}}

\mdfdefinestyle{questionstyle}{outerlinewidth = 0.2 , roundcorner = 2pt , leftmargin = 4 , rightmargin = 4 , backgroundcolor = gray!15 , linecolor = gray!25 , innertopmargin = 10pt , innerbottommargin = 10pt , everyline = true , splittopskip = 18pt , splitbottomskip = 10pt}
\newcommand{\question}[1]{
\begin{mdframed}[style=questionstyle]
\textbf{Exercise \thequestion} \quad #1
\end{mdframed}
\stepcounter{question}}

\ifanswers
\newcounter{answer}
\stepcounter{answer}
\mdfdefinestyle{answerstyle}{outerlinewidth = 0.2 , roundcorner = 2pt , leftmargin = 4 , rightmargin = 4 , backgroundcolor = gray!15 , linecolor = gray!25 , innertopmargin = 10pt , innerbottommargin = 10pt , everyline = true , splittopskip = 18pt , splitbottomskip = 10pt}
\newcommand{\answer}[1]{
\begin{mdframed}[style=answerstyle]
\textbf{Solution \theanswer} \quad #1
\end{mdframed}
\stepcounter{answer}}
\else
\newcommand{\answer}[1]{\ignorespaces}
\fi
    
\newcommand{\ba}{\begin{align}}

\newcommand{\be}{\begin{equation}}
\newcommand{\ee}{\end{equation}}

\newcommand{\ket}[1]{| #1 \rangle}
\newcommand{\bra}[1]{\langle  #1 |}

\allowdisplaybreaks[1]

\title{Superconformal symmetry and representations}

\author{Lorenz Eberhardt}
\affiliation{School of Natural Sciences, Institute for Advanced Study, \\
\hspace*{0.3cm}Princeton, NJ 08540, USA}
\emailAdd{elorenz@ias.edu}

\abstract{We give an introduction to conformal and superconformal algebras and their representations in various dimensions. Special emphasis is put on 4d $\mathcal{N}=2$ superconformal symmetry. This is the writeup of the lectures given at the Winter School ``YRISW 2020'' to appear in a special issue of JPhysA.}

\begin{document}
\maketitle

\makeatletter
\g@addto@macro\bfseries{\boldmath}
\makeatother

\section{Introduction}
Superconformal symmetry has played a major role in theoretical physics for the last 20 years. Conformal symmetry puts stringent consistency conditions on the theories and consequently CFTs are under much better control than normal QFTs. Combined with supersymmetry, this lets one solve some aspects of the theories completely.

Superconformal field theories (SCFTs) can be studied by a variety of methods and are ubiquitous in many physically relevant settings. Superconformal field theories can be effectively constrained by the conformal bootstrap
\cite{Rattazzi:2008pe}.
The AdS/CFT correspondence \cite{Maldacena:1997re} relates superconformal field theories to gravity theories on AdS. The holographic duals of many superconformal field theories are known, which gives insights in their strong-coupling behaviour.
There are many constructions of superconformal field theories that do not necessarily admit Lagrangian descriptions, such as class $\mathcal{S}$ theories \cite{Gaiotto:2009hg, Gaiotto:2009we} and Argyres-Douglas theories \cite{Argyres:1995jj}.
SCFTs are related through a web of dualities. These are mostly strong-weak dualities, such as S-duality in $\mathcal{N}=4$ SYM \cite{Seiberg:1994rs, Seiberg:1994aj, Argyres:2007cn}. SCFTs with enough supersymmetry admit certain protected subsectors, such as the chiral algebra in 4d $\mathcal{N}=2$ theories \cite{Beem:2013sza} or the 1d topological sector in 3d $\mathcal{N}=4$ SCFTs \cite{Chester:2014mea}. One can also define protected indices, which allows one to give strong evidence for various dualities \cite{Romelsberger:2005eg, Kinney:2005ej}.

Most of these results hinge on a good understanding of the relevant underlying symmetry algebras. In these introductory lectures, we briefly review conformal symmetry and its representations in arbitrary dimensions. We then move on to superconformal symmetry and discuss the case of 4d $\mathcal{N}=2$ in more detail. We follow the conventions of \cite{Cordova:2016emh}, although the representations of the 4d $\mathcal{N}=2$ superconformal algebra were already  completely understood by Dolan \& Osborn much earlier \cite{Dolan:2001tt, Dolan:2002zh}, see also \cite{Dobrev:1985qv}. Since these lectures are introductory, the list of references is by no means exhaustive. The Tables~\ref{tab:Q bounds N1}, \ref{tab:representations N1}, \ref{tab:Q bounds N2} and \ref{tab:representations N2} are directly taken from \cite{Cordova:2016emh}.
We have included some exercises and their solutions in the text.

\section{Conformal symmetry}\label{sec:conformal symmetry}
Conformal transformations are angle-preserving transformations. Equivalently, under conformal transformations, the metric can pick up at most a (position dependent) overall factor. A conformal transformation $\phi:\mathds{R}^{m,n} \longrightarrow \mathds{R}^{m,n}$ satisfies
\be 
\eta_{\mu\nu}\frac{\partial\phi^\mu(x)}{\partial x^\alpha} \frac{\partial \phi^\nu(x)}{\partial x^\beta}=\lambda(x)^2 \eta_{\alpha\beta}\ . \label{eq:conformal transformation}
\ee
Here, $\lambda(x)$ is the scale factor of the conformal transformation. $\eta_{\mu\nu}$ is the metric on $\mathds{R}^{m,n}$, having $m$ $+1$'s and $n$ $-1$'s on the diagonal. Let $d=m+n$ be the dimension of the space.
The conformal group in dimensions $d \ge 3$ consists of the following generators:
\begin{enumerate}
\item Lorentz transformations. A Lorentz transformation takes the form
\be 
\phi^\mu (x)=\tensor{\mathbf{M}}{^\mu_\nu} x^\nu\ ,
\ee
where $\tensor{\mathbf{M}}{^\mu_\nu}$ preserves the metric $\eta_{\mu\nu}$ and is hence an element of $\text{SO}(m,n)$.  Lorentz transformations are isometries and hence the scale factor $\lambda(x)=1$ is trivial.
\item Translations. A translation takes the form
\be 
\phi^\mu(x)=x^\mu+\mathbf{P}^\mu\ .
\ee
These are also isometries and hence the scale factor $\lambda(x)=1$ is trivial.
\item Dilatations. This is the first generator of the conformal group that it is not part of the Poincar\'e group. It acts by scaling
\be 
\phi^\mu(x)=\mathbf{D} x^\mu\ .
\ee
Correspondingly, the scale factor is $\lambda(x)=\mathbf{D}$.
\item Inversion. The inversion is a discrete generator. It acts by
\be 
\phi^\mu(x)=\frac{x^\mu}{|x|^2}\ .
\ee
The scale factor reads $\lambda(x)=|x|^{-2}$.
\item Special conformal transformations. Special conformal transformations can be understood as a composition of inversion $\circ$ translation $\circ$ inversion, which is then also conformal. They take the form
\be 
\phi^\mu(x)=\frac{x^\mu+|x|^2\mathbf{K}^\mu}{1+2x^\nu \mathbf{K}_\nu +|x|^2 |\mathbf{K}|^2}
\ee
The scale factor is
\be 
\lambda(x)=(1+2x^\mu \mathbf{K}_\mu +|x|^2 |\mathbf{K}|^2)^{-1}\ .
\ee
\end{enumerate}
\subsection{Dimensions \texorpdfstring{$d \ge 3$}{d>=3}}
For $d\ge 3$, this is a complete description of the conformal group. The group has dimension
\be 
\underbrace{\tfrac{1}{2}d(d-1)}_{\text{Lorentz}}+\underbrace{d}_{\text{translations}}+\underbrace{d}_{\text{special conformal transformations}}+\underbrace{1}_{\text{dilatation}}=\tfrac{1}{2}(d+2)(d+1)\ .
\ee
It is in fact isomorphic to the group $\mathrm{SO}(m+1,n+1)$. The inversion is an additional discrete generator which is not continuously connected to the identity (which turns the full group of conformal transformations into $\mathrm{O}(m+1,n+1)$). We will denote the corresponding generators of the Lie algebra $\mathfrak{so}(m+1,n+1)$ by $M_{\mu\nu}$, $P_\mu$, $D$ and $K_\mu$. They satisfy the following commutation relations:
\begin{subequations}
\begin{align}
[M_{\mu\nu},M_{\rho\sigma}] &= \eta_{\nu\rho}M_{\mu\sigma} + \eta_{\mu\sigma}M_{\nu\rho} - \eta_{\mu\rho}M_{\nu\sigma} - \eta_{\nu\sigma}M_{\mu\rho}\ , \\
[M_{\mu\nu},P_\rho] &= \eta_{\nu \rho} P_\mu-\eta_{\mu\rho} P_{\nu}  \ , \\
[M_{\mu\nu},K_\rho] &=  \eta_{\nu \rho} K_\mu-\eta_{\mu\rho} K_{\nu}  \ , \\
[D,P_\mu]&=P_\mu \ , \\
[D,K_\mu]&=-K_\mu \ , \\
[K_\mu,P_\nu]&=2(\eta_{\mu\nu}D-M_{\mu\nu}) \ .
\end{align} \label{eq:conformal algebra}
\end{subequations}
Unspecified commutators vanish.
It is important to understand the meaning of these commutation relations. The commutator $[M_{\mu\nu},M_{\rho\sigma}]$ defines the the Lorentz Lie algebra $\mathfrak{so}(m,n)$. The next two lines express the fact that both the translations $P_\mu$ and the special conformal transformation $K_\mu$ transform in the vector representation of $\mathfrak{so}(m,n)$. This was to be expected, since we specify translations and special conformal transformations by a vector, which has to transform covariantly under special conformal transformations. The next two lines specify how translations and special conformal transformations transform under the dilatation generator. $D$ defines the abelian Lie algebra $\mathds{R} \cong \mathfrak{so}(1,1)$ and hence every other generator has some definite weight under it.\footnote{In other words, the Cartan subalgebra of the conformal algebra consists of the Cartan subalgebra of the Lorentz algebra $\mathfrak{so}(m,n)$, together with the dilatation generator $D$.} This weight is called \emph{scaling dimension}. Translations have weight $+1$, whereas special conformal transformations have weight $-1$. Finally, the last line is needed for the closure of the algebra. It will play an important role in the following.

Note that the action of the translation generator on fields is given by a derivative. Hence, we will often identify $P_\mu=\partial_\mu$ with the derivative operator.
\subsection{Dimension \texorpdfstring{$d=2$}{d=2}}
The two-dimensional case is special. Let us consider Minkowski signature, $(m,n)=(1,1)$. The generators of the previous subsection are still generators, but there are infinitely many more generators and the conformal group is infinite-dimensional. To see this, it is convenient to introduce light-cone coordinates $x^+=\tfrac{1}{2}(x^0+x^1)$ and $x^-=\tfrac{1}{2}(x^0-x^1)$. In these coordinates, the metric takes the factorised form
\be 
\eta=-4\, \mathrm{d}x^+ \otimes \mathrm{d}x^-\ .
\ee 
Because of this, every transformation of the form
\be 
\phi^+(x)=\phi^+(x^+)\ , \qquad \phi^-(x)=\phi^-(x^-)
\ee
or 
\be 
\phi^+(x)=\phi^+(x^-)\ , \qquad \phi^-(x)=\phi^-(x^+)
\ee
is conformal. Hence there is an entire functions worth of conformal transformations. They form the group $\mathrm{Diff}(\mathds{R}) \times \mathrm{Diff}(\mathds{R})$, since we can choose any pair of diffeomorphisms in the two light-cone directions.\footnote{Because $\phi^+$ can be either a function of $x^+$ or $x^-$, the conformal group is actually a $\mathds{Z}_2$-extension $(\mathrm{Diff}(\mathds{R}) \times \mathrm{Diff}(\mathds{R})) \rtimes \mathds{Z}_2$, where $\mathds{Z}_2$ permutes the two copies.}  The global group $\mathrm{SO}(2,2)$ we discussed above is the subgroup (up to discrete identifications)
\be 
\mathrm{SO}(2,2)\cong \mathrm{SL}(2,\mathds{R})\times \mathrm{SL}(2,\mathds{R})\subset \mathrm{Diff}(\mathds{R}) \times \mathrm{Diff}(\mathds{R})
\ee
of fractional linear transformations on the two light-cone coordinates. For Euclidean signature, we have $\mathrm{SO}(3,1) \cong \mathrm{SL}(2,\mathds{C})$ (again up to global identifications). $\mathrm{SL}(2,\mathds{C})$ should be thought of as fractional linear transformations on the complex plane $\mathds{C}$. The full conformal group actually consists of  \emph{all} (anti)holomorphic maps. For more details on the 2-dimensional case, we refer to \cite{DiFrancesco:1997nk, Blumenhagen:2009zz}.

\question{\textbf{Embedding space formalism.}  \questionnumber{embedding space} An elegant way to formulate the action of the conformal group is via the so-called embedding space formalism. For this, we embed the physical space $\mathds{R}^{m,n}$  with $d=m+n$ into the null-light cone of $\mathds{R}^{m+1,n+1}$ as follows:
\be 
P^0=\frac{1+|x|^2}{2}\ , \qquad P^\mu=x^\mu\ , \qquad P^{d+1}=\frac{1-|x|^2}{2}\ .
\ee
We take the norm in $\mathds{R}^{m+1,n+1}$ to be
\be 
-(P^0)^2+P^\mu P^\nu \eta_{\mu \nu}+(P^{d+1})^2\ , \label{eq:embedding space metric}
\ee
where $\eta_{\mu\nu}$ is the original metric, which makes the image of $\mathds{R}^{m,n}$ inside $\mathds{R}^{m+1,n+1}$ null.  Thus, the image is indeed null. We can furthermore identify $P^\mu \sim \lambda P^\mu$ and hence obtain an embedding of $\mathbb{R}^{m,n}$ into the projectivization of the null-cone.

Show that conformal transformations act linearly on $P^\mu$,
\be 
P \longmapsto A P
\ee
for some matrix $A \in \mathrm{SO}(m+1,n+1)$. Thus, the embedding space makes the action of the conformal group manifest and it is thus often useful to work with the coordinates $P$ instead.
}
\answer{We check this for the generators of the conformal group separately.
\begin{enumerate}
\item Lorentz transformations. This is obvious, since $P^0$ and $P^{d+1}$ are untouched by Lorentz transformations, they act linearly on the embedding space coordinates.
\item Translations. $P$ transforms as
\begin{align}
P &\mapsto \left(\frac{1+|x+a|^2}{2},x^\mu+a^\mu,\frac{1-|x+a|^2}{2}\right) \\
&=\begin{pmatrix}
1+\frac{|a|^2}{2} & a_1 & \cdots & a_d & \frac{|a|^2}{2} \\
a^1 & 1 & \cdots & 0 & a^1\\
\vdots & \vdots & \ddots & \vdots & \vdots\\
a^d & 0 & \cdots & 1 & a^d\\
-\frac{|a|^2}{2} & -a_1 & \cdots & -a_d & 1-\frac{|a|^2}{2}
\end{pmatrix}P\ .
\end{align}
It is then straightforward to see that the matrix is an element of $\mathrm{SO}(m,n)$.
\item Inversion. $P$ transforms as
\begin{align}
P &\mapsto \frac{1}{|x|^2}\left(\frac{1+|x|^2}{2}\ , x^\mu,-\frac{1-|x|^2}{2}\right)\ , \\
&\sim \mathrm{diag}(1,\dots,1,-1) P \ ,
\end{align}
where we used that we identified multiples of $P^\mu$. Note that this matrix is actually only in $\mathrm{O}(m,n)$ and is hence not continuously connected to the identity.
\end{enumerate}
Since compositions of these generators produce the entire conformal group, we are done.
}

\section{Unitary representations of the conformal group}
Next, we study unitary representations of the conformal group. We will work from now on entirely in Euclidean signature (i.e.~setting $n=0$ above). We will also assume $d \ge 3$. A good reference on this material is \cite{Rychkov:2016iqz}. For $d=2$, the discussion is still true for the subgroup of global conformal transformations, but representations of the full conformal group (the Virasoro algebra) are more complicated \cite{DiFrancesco:1997nk}.
\subsection{Highest weight representations}
The dilatation operator $D$ can be identified with the Hamiltonian in this context. Euclidean CFTs are usually quantised in radial quantisation on the plane. In this quantisation scheme, time translations are given by dilatation. Hence, we will always diagonalise $D$ in a given representation. The eigenvalue $\Delta$ is called the \emph{scaling dimension}.

Any physical representation of the conformal group should have bounded energy spectrum from below. Now recall that translation generators $P_\mu$ have scaling dimension $+1$, whereas special conformal transformation generators $K_\mu$ have scaling dimension $-1$. Since there has to be one state that has lowest energy, there exists a highest weight state
\be 
K_\mu \ket{[L]_\Delta}=0\ .
\ee
This highest weight state is also called \emph{primary state}. 
This primary state has the label $\Delta$, which is its scaling dimension. Moreover, it still forms a representation with respect to the Lorentz group $\mathfrak{so}(d)$. So to get a unique primary state, one should also require it to be a highest weight state of the $\mathfrak{so}(d)$ representation. The index $L$ labels the respective representation of the Lorentz group. We will later discuss mostly the case $d=4$, where $\mathfrak{so}(4) \cong \mathfrak{su}(2)\oplus \mathfrak{su}(2)$ and hence representations are  specified by two $\mathfrak{su}(2)$ spins $[L]=[j;\bar{\jmath}]$.

The corresponding fields are called \emph{primary fields}. Their transformation behaviour under a conformal transformation $\phi$ is
\be 
(\phi \cdot \Phi)(x)=\lambda(x')^{-\Delta}(\rho(R(x')) \cdot\Phi)(x') \ , \label{eq:primary field transformation}
\ee
where $x'=\phi^{-1}(x)$.
Here, $\lambda(x')$ is the scale factor associated to a conformal transformation, see eq.~\eqref{eq:conformal transformation} and $R(x')$ is the rotational part of the conformal transformation. $\rho$ is then the $\mathrm{SO}(d)$ representation as specified by $[j;\bar{\jmath}]$.

Let us return to the action of the conformal algebra on the Hilbert space of the theory. The non-highest weight states in the representation are called \emph{descendants} and they are obtained by acting with the translation operator $P_\mu$ and Lorentz generators on the highest weight state:\footnote{It can happen that a state is both of this form and again a primary state, in which case the state becomes null and decouples from the theory. We will shortly see examples of this.}
\be 
P_{\mu_1}\cdots P_{\mu_n}\ket{[L]_\Delta}\ .
\ee
\question{\textbf{Scalar three-point function.} \questionnumber{scalar three point function} In this exercise, we compute a three-point function
\be 
\langle \Phi_1(x_1) \Phi_2(x_2) \Phi_3(x_3) \rangle
\ee
of primary scalar fields. Show that conformal symmetry restricts it to have the form
\begin{multline} 
\langle \Phi_1(x_1) \Phi_2(x_2) \Phi_3(x_3) \rangle\\
=\frac{C_{123}}{|x_1-x_2|^{\Delta_1+\Delta_2-\Delta_3}|x_1-x_3|^{\Delta_1+\Delta_3-\Delta_2}|x_2-x_3|^{\Delta_2+\Delta_3-\Delta_1}}
\end{multline}
for some constant $C_{123}$.
}
\answer{This computation is done most efficiently in the embedding space formalism. We can view $\Phi(P)$ to be a field in the embedding space introduced in Exercise~\ref{embedding space}. The answer has to be an $\mathrm{SO}(d+1,1)$ singlet, which imposes that
\be 
\langle \Phi_1(P_1) \Phi_2(P_2) \Phi_3(P_3) \rangle=f(P_1 \cdot P_2,P_2 \cdot P_3,P_3 \cdot P_1)\ ,
\ee
for some function $f$. The inner product is taken w.r.t.~the metric in the embedding space as specified by eq.~\eqref{eq:embedding space metric}.
In the embedding space, the scaling dimension is the homogeneity degree under the rescaling $P \mapsto \lambda P$, see eq.~\eqref{eq:primary field transformation}
\be 
\Phi(\lambda P)=\lambda^{-\Delta} \Phi(P)\ .
\ee
Thus, the correlator has to be homogeneous of degrees $-\Delta_1$, $-\Delta_2$ and $-\Delta_3$, which fixes it to have the form
\begin{multline}
\langle \Phi_1(P_1) \Phi_2(P_2) \Phi_3(P_3) \rangle\\
=\frac{C_{123}}{(-2P_1\cdot P_2)^{\frac{\Delta_1+\Delta_2-\Delta_3}{2}}(-2P_1\cdot P_3)^{\frac{\Delta_1+\Delta_3-\Delta_2}{2}}(-2P_2\cdot P_3)^{\frac{\Delta_2+\Delta_3-\Delta_1}{2}}}\ .
\end{multline}
The result follows from noting that $-2P_i \cdot P_j=|x_i-x_j|^2$.

Alternatively, one can work directly in position space, in which case Poincar\'e invariance imposes
\be 
\langle \Phi_1(x_1)\Phi_2(x_2)\Phi_3(x_3) \rangle=f(|x_1-x_2|,|x_1-x_3|,|x_2-x_3|)\ ,
\ee
for some function $f$. 
We then demand invariance under inversion, which ensures invariance under the whole conformal group. For this, we need
\begin{multline} 
\left\langle \Phi_1\left(\frac{x_1}{|x_1|^2}\right)\Phi_2\left(\frac{x_2}{|x_2|^2}\right)\Phi_3\left(\frac{x_3}{|x_3|^2}\right) \right\rangle\\
=|x_1|^{2\Delta_1}|x_2|^{2\Delta_2}|x_3|^{2\Delta_3}\langle \Phi_1(x_1)\Phi_2( x_2)\Phi_3( x_3) \rangle\ .
\end{multline}
We have
\be 
|x_i-x_j|^2\mapsto \frac{\big|x_i |x_j|^2-x_j|x_i|^2\big|^2}{|x_i|^4|x_j|^4}=\frac{|x_i-x_j|^2}{|x_i|^2|x_j|^2}\ ,
\ee
so combining this with the previous constraints gives the following functional equation for $f$:
\begin{multline} 
f\left(\frac{|x_1-x_2|^2}{|x_1|^2|x_2|^2},\frac{|x_1-x_3|^2}{|x_1|^2|x_3|^2},\frac{|x_2-x_3|^2}{|x_2|^2|x_3|^2}\right)\\
=|x_1|^{2\Delta_1}|x_2|^{2\Delta_2}|x_3|^{2\Delta_3} f(|x_1-x_2|,|x_1-x_3|,|x_2-x_3|)\ .
\end{multline}
Since the products $|x_i|^2|x_j|^2$ are all independent, this implies that $f$ has to be a homogeneous function of degree $-\Delta_i-\Delta_j+\Delta_k$ (where $k$ is the third index) in the respective slot. Thus, the result follows.
}
\subsection{Unitarity}
We will exclusively consider unitary theories, where all the states have non-negative norm. To talk about the norm, we have to introduce the hermitian conjugate. One way of motivating our definition of the hermitian conjugate is the following. We are working in Euclidean space-time, where time is complex and hermitian conjugation sends $t \to -t$. We have identified time with the radial direction in the plane (by our choice of Hamiltonian). This means that hermitian conjugation acts on the plane by the inversion element of the conformal group. Hence, hermitian conjugation should conjugate the conformal generators by the inversion element. For the Lorentz group, we use the usual hermitian conjugate, which is obtained from taking the transpose. This means that
\begin{subequations}
\begin{align}
M_{\mu\nu}^\dag&=M_{\nu\mu}=-M_{\mu\nu}\ , \\
P_\mu^\dag&=K_\mu\ , \\
K_\mu^\dag&=P_\mu\ , \\
D^\dag&=D\ .
\end{align}
\end{subequations}
With these preparations, we are ready to study the norms in representations of the conformal group. 
\paragraph{Scalars.}
Let us start with a scalar, i.e.$\,[L]=[0]$, the trivial representation. We can declare the highest weight state to have unit norm $\langle [0]_\Delta|[0]_\Delta \rangle=1$. We can then compute the norm of a general level one descendant:
\begin{align}
\lVert a^\mu P_\mu \ket{[0]_\Delta} \rVert^2&= \bar{a}^\mu a^\nu \bra{[0]_\Delta} K_\mu P_\nu \ket{[0]_\Delta}\\
&= \bar{a}^\mu a^\nu \bra{[0]_\Delta} [K_\mu, P_\nu] \ket{[0]_\Delta} \\
&= 2\bar{a}^\mu a^\nu \bra{[0]_\Delta} \delta_{\mu\nu} D-M_{\mu\nu} \ket{[0]_\Delta} \\
&=2|a|^2 \Delta\ . \label{eq:scalar level one norm}
\end{align}
Thus, we conclude that all level one states have positive norm, provided that $\Delta \ge 0$. $\Delta=0$ is an allowed value, since zero-norm states will simply decouple from the theory.

Next, we compute the norm at the second level. Let $a^{\mu\nu}$ be a symmetric 2-tensor.
\begin{align}
\lVert a^{\mu\nu} P_\mu P_\nu \ket{[0]_\Delta} \rVert^2&= a^{\mu\nu} \bar{a}^{\rho\sigma} \bra{[0]_\Delta} K_\sigma K_\rho P_\mu P_\nu \ket{[0]_\Delta}\\
&= 2a^{\mu\nu} \bar{a}^{\rho\sigma} \big(\bra{[0]_\Delta} K_\sigma (\delta_{\mu\rho}D+M_{\mu\rho}) P_\nu \ket{[0]_\Delta}\nonumber\\
&\qquad\qquad\qquad\qquad+\bra{[0]_\Delta} K_\sigma P_\mu (\delta_{\nu\rho}D+M_{\nu\rho}) \ket{[0]_\Delta}\big)\\
&= 2a^{\mu\nu} \bar{a}^{\rho\sigma} \big(2\delta_{\mu\rho}\delta_{\nu\sigma} \Delta(\Delta+1)+\bra{[0]_\Delta} K_\sigma (\delta_{\nu\rho}P_\mu-\delta_{\mu\nu} P_\rho) \ket{[0]_\Delta}\nonumber\\
&\qquad\qquad\qquad\qquad+2\Delta^2 \delta_{\mu\sigma}\delta_{\nu \rho}\big)\\
&= 4a^{\mu\nu} \bar{a}^{\rho\sigma} \big(\delta_{\mu\rho}\delta_{\nu\sigma} \Delta(\Delta+1)+\Delta(\Delta+1) \delta_{\mu\sigma}\delta_{\nu \rho}-\Delta\delta_{\mu\nu}\delta_{\rho\sigma}\big)\\
&=4\Delta(2(\Delta+1)\bar{a}^{\mu\nu}a_{\mu\nu}-\tensor{\bar{a}}{^\mu_\mu}\tensor{a}{^\nu_\nu})\ . \label{eq:scalar level 2 norm}
\end{align}
Let us analyse when this is positive. First, let us set $a^{\mu\nu}=\delta^{\mu\nu}$. Then the norm is easily evaluated and gives
\be 
4\Delta(2(\Delta+1) d-d^2)\ ,
\ee
which for $\Delta \ge 0$ is non-negative only if
\be 
\Delta\ge \frac{d}{2}-1\qquad\text{or}\qquad \Delta=0\ . \label{eq:scalar unitarity bound}
\ee
Hence we find a unitarity bound in this case. This bound is actually sufficient, since the norm of the traceless part $b^{\mu\nu}=a^{\mu\nu}-\tfrac{1}{d} \delta^{\mu\nu}\tensor{a}{^\rho_\rho}$ is
\be 
\bar{b}^{\mu\nu}b_{\mu\nu}=\bar{a}^{\mu\nu}a_{\mu\nu}-\frac{1}{d} \tensor{\bar{a}}{^\mu_\mu}\tensor{a}{^\nu_\nu}\ge 0\ .
\ee
One can similarly compute norms at higher level in the Verma module, but one does not find any stronger constraint than the unitarity bound \eqref{eq:scalar unitarity bound}. Proofs of this fact are actually quite nontrivial \cite{Mack:1975je, Yamazaki:2016vqi}. 

\question{\textbf{Unitarity bound for Vectors.} \questionnumber{unitarity bound vector} Repeat the same calculation for the vector representation $[L]=[V]$ and show that the level one norms impose the following unitarity bound:
\be 
\Delta \ge d-1\ .\label{eq:vector unitarity bound}
\ee
}
\answer{
Since the vector representation carries one index $\mu$, it is convenient to put an index on the highest weight state. We again compute the level one norm. We have for a 2-tensor $a^{\mu\nu}$
\begin{align}
\lVert a^{\mu\nu} P_\mu \ket{[V]_\Delta}_\nu \rVert^2&=a^{\mu\nu}\bar{a}^{\rho\sigma}\,{}_\sigma\bra{[V]_\Delta}K_\rho P_\mu \ket{[V]_\Delta}_\nu \\
&=2a^{\mu\nu}\bar{a}^{\rho\sigma}\, {}_\sigma\bra{[V]_\Delta}D \delta_{\mu\rho}+M_{\mu\rho} \ket{[V]_\Delta}_\nu \\
&=2a^{\mu\nu}\bar{a}^{\rho\sigma}\big((\Delta \delta_{\mu \rho}\, {}_\sigma\langle [V]_\Delta | [V]_\Delta \rangle_\nu +\delta_{\nu\rho}\, {}_\sigma\langle [V]_\Delta | [V]_\Delta \rangle_\mu\nonumber\\
&\qquad\qquad\qquad\qquad	-\delta_{\mu\nu}\, {}_\sigma\langle [V]_\Delta | [V]_\Delta \rangle_\rho\big)\\
&=2a^{\mu\nu}\bar{a}^{\rho \sigma} \big(\Delta \delta_{\mu\rho} \delta_{\nu \sigma}+\delta_{\nu\rho}\delta_{\mu\sigma}-\delta_{\mu\nu} \delta_{\sigma\rho}\big) \\
&=2\big(\Delta \bar{a}^{\mu\nu}a_{\mu\nu}+\bar{a}^{\mu\nu}a_{\nu\mu}-\tensor{\bar{a}}{^\mu_\mu}\tensor{a}{^\nu_\nu}\big)\ .
\end{align}
This has the same index structure as the norm for the scalar field at level 2, except that $a_{\mu\nu}$ is now not a symmetric tensor. Putting $a_{\mu\nu}= \delta_{\mu\nu}$ shows that $\Delta \ge d-1$ is a necessary condition for unitarity.
Consider now
\be 
b_{\mu\nu}=\frac{1}{2}\big(\sqrt{d}+\sqrt{d-2}\big)a_{\mu\nu}+\frac{1}{2}\big(\sqrt{d}-\sqrt{d-2}\big)a_{\nu\mu}-\frac{1}{\sqrt{d}} \tensor{a}{^\rho_\rho}\delta_{\mu\nu}\ .
\ee
Positivity of the norm of $b_{\mu\nu}$ gives
\be 
\bar{b}^{\mu\nu}b_{\mu\nu}=(d-1)\bar{a}^{\mu\nu}a_{\mu\nu}+\bar{a}^{\mu\nu}a_{\nu\mu}-\tensor{\bar{a}}{^\mu_\mu}\tensor{a}{^\nu_\nu}\ge 0\ ,
\ee
which shows that for $\Delta \ge d-1$, the norm of the level 1 descendants is always positive, regardless of the choice of $a_{\mu\nu}$.
There are no more conditions from evaluating norms at higher levels.
}

\paragraph{General case.} Let us discuss the general case. One can use representation theory to determine the norm of the level 1 states. Let us denote indices of the representation $L$ by $\alpha$, $\beta$, \dots We want to evaluate the eigenvalues of the operator
\begin{align}
{}_{\alpha}\langle [L]_\Delta| M_{\mu\nu} | [L]_\Delta \rangle_\beta\ \label{eq:M expectation value}
\end{align}
To do so, one can use the following trick:
\be 
M_{\mu\nu}=\frac{1}{2}(\delta_{\mu \rho}\delta_{\nu \sigma}-\delta_{\mu \sigma}\delta_{\nu \rho}) M_{\rho \sigma}= (V \cdot M)_{\mu\nu}\ ,
\ee
where we used the generators of the vector representation of $\mathfrak{so}(d)$. The inner product for arbitrary representations is defined as $A \cdot B=\tfrac{1}{2} A_{\mu\nu}B_{\mu\nu}$ (and is still valued in the tensor product space of the representations $A$ and $B$). We can then write
\be 
V \cdot M=\frac{1}{2}\left((V+M) \cdot (V+M)-V \cdot V- M \cdot M \right)\ .
\ee
The quadratic terms are precisely the quadratic Casimirs of the respective representations. Thus, the eigenvalues of the operator \eqref{eq:M expectation value} are
\be 
\frac{1}{2}\left(\mathcal{C}(L')-\mathcal{C}(V)-\mathcal{C}(L)\right)\ ,
\ee
where $L'$ runs over all representations in the tensor product $L \otimes V$. Positivity of the norm at this level is hence the statement 
\be 
\forall L' \in L \otimes V\,:\qquad \Delta \ge \frac{1}{2}\left(\mathcal{C}(V)+\mathcal{C}(L)-\mathcal{C}(L')\right)\ .
\ee
It turns out that only for the scalar, there is another constraint coming from the norm at level 2. Let us list the quadratic Casimirs for some important representations:
\begin{subequations}
\begin{align}
\mathcal{C}(\text{scalar})&=0\ , \\
\mathcal{C}(\text{vector})&=d-1\ ,\\
\mathcal{C}(\text{symmetric traceless 2-tensor})&=2d\ , \\
\mathcal{C}(\text{anti-symmetric 2-tensor})&=2d-4\ , \\
\mathcal{C}(\text{spinor})&=\frac{d}{8}(d-1)\ , \\
\mathcal{C}(\text{spin $\tfrac{3}{2}$})&=\frac{d}{8}(d+7)\ .
\end{align}
\end{subequations}
They lead respectively to the unitarity bounds
\begin{subequations}
\begin{align}
\text{vector}&: &\Delta &\ge d-1\ , & \text{for } L'&=\text{scalar}\ , \\
\text{symmetric traceless 2-tensor}&: &\Delta &\ge d\ , & \text{for } L'&=\text{vector}\ , \\
\text{anti-symmetric 2-tensor}&:&\Delta &\ge d-2\ , & \text{for } L'&=\text{vector}\ , \\
\text{spinor}&:&\Delta &\ge \frac{d-1}{2}\ , & \text{for } L'&=\text{(conjugate) spinor}\ , \\
\text{spin $\tfrac{3}{2}$}&:&\Delta &\ge d-\frac{1}{2}\ , & \text{for } L'&=\text{spinor}\ .
\end{align}
\end{subequations}
When the unitarity bounds are saturated, there are null-vectors (i.e.~vectors with zero norm) in the representation. Hence representations with these null-vectors always have to saturate the bound and cannot acquire any anomalous dimensions in the quantum theory.
\paragraph{Four dimensions.} Let us discuss in more detail the case $d=4$. In this case, $\mathfrak{so}(4) \cong\mathfrak{su}(2) \oplus \mathfrak{su}(2)$ and hence Lorentz representations are labelled by two spins $[j;\bar{\jmath}]$. We will take $j$ and $\bar{\jmath}$ to be Dynkin labels, i.e.~the spin $s$ representation has $j=2s$. If $j>0$ and $\bar{\jmath}>0$, we can always choose $L'=[j-1;\bar{\jmath}-1]$. If however $j=0$, we can only choose $L'=[j+1;\bar{\jmath}-1]$ and similarly for $\bar{\jmath}$. The Casimir for an $\mathfrak{su}(2)$ representation in this normalisation reads $\mathcal{C}(j)=\tfrac{1}{2}j(j+2)$. This leads to the unitarity bound in four dimensions
\be 
\Delta \ge \begin{cases}
f(j)+f(\bar{\jmath})\ , \quad &j>0\ \text{or}\ \bar{\jmath}>0\ , \\
1\ , \quad &j=\bar{\jmath}=0\ . \label{eq:unitarity bound 4d bosonic}
\end{cases}
\ee 
with $f(j)=\tfrac{j}{2}+1$ if $j>0$ and $f(0)=0$.
\question{\textbf{Recombination rules.}  \questionnumber{recombination rules}Usually, conformal field theories depend on several parameters, such as the coupling constants of the theory. As these parameters are tuned, the scaling dimension of the multiplets typically changes, but the total number of states typically does not change. Thus, we have the phenomenon of recombination: as we move around in the parameter space of the theory, two or more short multiplets may join up to form one long multiplet, whose dimension is no longer protected by the unitarity bound. 
Argue the following recombination rule:
\begin{align}
[V]_{\Delta=d-1} \oplus [0]_{\Delta=d} \longrightarrow [V]_{\Delta=d-1+\varepsilon}\ , \label{eq:recombination rule vector}
\end{align}
where the multiplets on the right do not have a null-descendant anymore and can hence leave the unitarity bound, which we denoted by a shift of the scaling dimension by $\varepsilon>0$.
}
\answer{The representation $[V]_{\Delta=d-1} $ has the null-descendant $P^\mu |[V]_{d-1}\rangle_\mu$ at level 1, where we use the same notation as in the solution of Exercise~\ref{unitarity bound vector}. This descendant is actually again a primary field, since
\be 
K_\nu P^\mu |[V]_{d-1}\rangle_\mu\propto |[V]_{d-1}\rangle_\nu\ ,
\ee
by comparing scaling dimensions on both sides. But as we checked in Exercise~\ref{unitarity bound vector}, the norm vanishes and hence $K_\nu P^\mu |[V]_{d-1}\rangle_\mu=0$. This means that the second term of the left-hand side fills the gap in the representation left by the null-descendant. As we stated above, no new primary fields appear at higher level. Thus, when we move away from the unitarity bound, we need another scalar primary that fills the gap and to complete the multiplet.
}
\subsection{Examples}
\paragraph{The identity field.} Let us start with a trivial example. Every CFT has an identity field $\mathds{1}$, which has weight $\Delta=0$ and is a scalar. Hence we know from our study above that all its descendants have to be null, see eqs.~\eqref{eq:scalar level one norm} and \eqref{eq:scalar level 2 norm}, which are proportional to $\Delta$. One can easily convince oneself that this continues to hold true for arbitrary levels. This is indeed the case, since the identity is translation-invariant,
\be 
P_\mu \mathds{1}=\partial_\mu \mathds{1}=0\ .
\ee
\paragraph{Free scalar.} The theory of a massless free scalar field is a CFT. It has Lagrangian
\be 
\int \mathrm{d}^d x \ \partial_\mu \phi \partial^\mu \phi\ .
\ee
Since the action is dimensionless, we conclude that $\phi$ has to have dimension $\tfrac{d}{2}-1$ (a derivative has dimension 1 and $\mathrm{d}x$ has dimension $-1$).  Since the theory is free, this is not corrected at the quantum level. This exactly saturates the bound for the scalar. Hence we expect there to be a null-vector in the representation of the scalar field. This is indeed the case, since we have
\be 
P_\mu P^\mu \phi=\partial_\mu \partial^\mu \phi=0\ .
\ee
\paragraph{Free fermion.} The theory of a massless (Dirac) fermion is also conformal with Lagrangian
\be 
i \int \mathrm{d}^d x \ \overline{\psi} \gamma^\mu \partial_\mu  \psi\ .
\ee
By the same reasoning as above, the free fermion has dimension $\Delta=\tfrac{d-1}{2}$, saturating the unitarity bound we derived above. Thus we expect again that there is a null-vector in the representation. Moreover,we expect to see it at level 1, since we derived the unitarity bound above at level 1. This is again realised thanks to the equations of motion,
\be 
\gamma^\mu P_\mu \psi=\gamma^\mu \partial_\mu \psi=0\ .
\ee
\paragraph{Conserved currents.} A symmetry of the theory gives rise to conserved currents. The conservation equation of a current $j^\mu$ reads
\be 
P_\mu j^\mu=\partial_\mu j^\mu=0
\ee
and hence the current $j^\mu$ is part of a conformal representation with a null-vector. Since $j^\mu$ transforms tensorially under coordinate transformations, one can also conclude that $j^\mu$ is a primary field. Thus, the conserved current has to saturate the unitarity bound above, so in a CFT we have always
\be 
\Delta(j_\mu)=d-1\ ,
\ee
without any quantum corrections. This is also the natural dimension for a conserved current, since the associated conserved charge
\be 
Q=\int_\text{space} \mathrm{d}^{d-1}x\ j^0
\ee
should be dimensionless.
\paragraph{Stress-energy tensor.} Any QFT possesses a stress-energy tensor. It is a symmetric tensor and for CFTs traceless. The stress-energy tensor $T_{\mu\nu}$ is moreover conserved,
\be 
P_\mu T^{\mu\nu}=0\ ,
\ee
and hence has to saturate by the above reasoning the unitarity bound
\be 
\Delta(T^{\mu\nu})=d\ .
\ee
\paragraph{Supercurrents.} Below, we will also encounter the supersymmetric partner of the stress-energy tensor, the so-called supercurrents. They have spin $\tfrac{3}{2}$ and are conserved and hence have scaling dimension
\be 
\Delta(G^{\mu\alpha})=d-\frac{1}{2}\ .
\ee

\section{Superconformal symmetry} \label{sec:superconformal symmetry}
In this section, we discuss the various superconformal algebras, focussing mostly on four dimensions. Standard references for this material are \cite{Dobrev:1985qv, Minwalla:1997ka, Dolan:2001tt, Dolan:2002zh, Kinney:2005ej, Cordova:2016emh}.
\subsection{Four dimensions}
To simplify the discussion, we will complexify the conformal algebra. Reality conditions will be imposed later by specifying the hermitian conjugate in the algebra.

In four dimensions, the Lorentz group is $\mathfrak{so}(4,\mathds{C}) \cong \mathfrak{sl}(2,\mathds{C}) \oplus \mathfrak{sl}(2,\mathds{C})$. Similarly, the conformal group is $\mathfrak{so}(6,\mathds{C}) \cong \mathfrak{sl}(4,\mathds{C})$. Because of this, it is very useful to employ a bispinor notation for the conformal algebra. We denote
\begin{subequations}
\begin{align}
(\sigma^\mu)_{\alpha\dot{\alpha}}&=(i\mathds{1}, \, \vec{\boldsymbol{\sigma}})\ , \\
(\bar{\sigma}^\mu)^{\dot{\alpha}\alpha}&=(-i\mathds{1}, \, \vec{\boldsymbol{\sigma}})\ ,
\end{align}
\end{subequations}
where $\vec{\boldsymbol{\sigma}}$ is the vector of Pauli matrices. $\mathfrak{su}(2)$ indices are lowered and raised by
\be 
X^a=\epsilon^{ab}X_b\ , \qquad X_a=\epsilon_{ab}X^b\ ,
\ee
where $\epsilon_{12}=1$, $\epsilon^{12}=-1$.

The Lorentz generators $M_{\mu\nu}$ split into two adjoint representations of $\mathfrak{sl}(2,\mathds{C})$, which we will denote by $\tensor{M}{_\alpha^\beta}$ and $\tensor{M}{^{\dot{\alpha}}_{\dot{\beta}}}$, respectively. The momentum and special conformal transformations are bispinors under the Lorentz group. We write 
\begin{subequations}
\begin{align}
P_{\alpha\dot{\alpha}}&=\tensor{(\sigma^\mu)}{_{\alpha\dot{\alpha}}} P_\mu\ , \\
K^{\dot{\alpha}\alpha}&=\tensor{(\bar{\sigma}^\mu)}{^{\dot{\alpha}\alpha}} K_\mu\ , \\
\tensor{M}{_\alpha^\beta}&=-\tfrac{1}{4} (\bar{\sigma}^\mu)^{\dot{\alpha}\beta} (\sigma_\nu)_{\alpha\dot{\alpha}} M_{\mu\nu}\ , \\
\tensor{M}{^{\dot{\alpha}}_{\dot{\beta}}}&=-\tfrac{1}{4} (\bar{\sigma}^\mu)^{\dot{\alpha}\alpha} (\sigma_\nu)_{\alpha\dot{\beta}} M_{\mu\nu}\ .
\end{align}
\end{subequations}
The conformal algebra reads in this notation, see e.g.~\cite[Appendix A]{Bianchi:2019sxz}
\begin{subequations}
\begin{align}
[\tensor{M}{_\alpha^\beta},\tensor{M}{_\gamma^\delta}]&=\tensor{\delta}{_\gamma^\beta} \tensor{M}{_\alpha^\delta}-\tensor{\delta}{_\alpha^\delta} \tensor{M}{_\gamma^\beta}\ ,\\
[\tensor{M}{^{\dot{\alpha}}_{\dot{\beta}}},\tensor{M}{^{\dot{\gamma}}_{\dot{\delta}}}]&=-\tensor{\delta}{^{\dot{\alpha}}_{\dot{\delta}}} \tensor{M}{^{\dot{\gamma}}_{\dot{\beta}}}+\tensor{\delta}{^{\dot{\gamma}}_{\dot{\beta}}} \tensor{M}{^{\dot{\alpha}}_{\dot{\delta}}}\ ,\\
[\tensor{M}{_\alpha^\beta},\tensor{P}{_{\gamma\dot{\gamma}}}]&=\tensor{\delta}{_\gamma^\beta} \tensor{P}{_{\alpha\dot{\gamma}}}-\tfrac{1}{2}\tensor{\delta}{_\alpha^\beta}P_{\gamma\dot{\gamma}}\ ,\\
[\tensor{M}{^{\dot{\alpha}}^{\dot{\beta}}},\tensor{P}{_{\gamma\dot{\gamma}}}]&=\tensor{\delta}{^{\dot{\alpha}}_{\dot{\gamma}}} \tensor{P}{_{\gamma\dot{\beta}}}-\tfrac{1}{2}\tensor{\delta}{^{\dot{\alpha}}_{\dot{\beta}}}P_{\gamma\dot{\gamma}}\ ,\\
[\tensor{M}{_\alpha^\beta},\tensor{K}{^{\dot{\gamma}\gamma}}]&=-\tensor{\delta}{_\alpha^\gamma} \tensor{K}{^{\dot{\gamma}\beta}}+\tfrac{1}{2}\tensor{\delta}{_\alpha^\beta}K^{\dot{\gamma}\gamma}\ ,\\
[\tensor{M}{_\alpha^\beta},\tensor{K}{^{\dot{\gamma}\gamma}}]&=-\tensor{\delta}{^{\dot{\gamma}}_{\dot{\beta}}} \tensor{K}{^{\dot{\alpha}\gamma}}+\tfrac{1}{2}\tensor{\delta}{^{\dot{\alpha}}_{\dot{\beta}}}K^{\dot{\gamma}\gamma}\ ,\\
[D,P_{\alpha\dot{\alpha}}]&=P_{\alpha\dot{\alpha}}\ , \\
[D,K^{\dot{\alpha}\alpha}]&=-K^{\dot{\alpha}\alpha}\ , \\
[K^{\dot{\alpha}\alpha},P_{\beta\dot{\beta}}]&=4 \tensor{\delta}{_\beta^\alpha}\tensor{\delta}{^{\dot{\alpha}}_{\dot{\beta}}} D+4 \tensor{\delta}{_\beta^\alpha}\tensor{M}{^{\dot{\alpha}}_{\dot{\beta}}} +4 \tensor{\delta}{^{\dot{\alpha}}_{\dot{\beta}}} \tensor{M}{_\beta^\alpha}\ .
\end{align}
\end{subequations}
We now want to introduce supercharges $\tensor{Q}{^i_\alpha}$ and $\tensor{\bar{Q}}{_{i\dot{\alpha}}}$. Here, $i=1$, \dots, $\mathcal{N}$ is the amount of supersymmetry we want to consider. $\mathcal{N}=1$ is the minimal amount and $\mathcal{N}=4$ is the maximal amount. The algebra also has an R-symmetry $\mathfrak{u}(\mathcal{N})$, which acts on the indices $i$ and $j$. The case $\mathcal{N}=4$ will be slightly exceptional.
The R-symmetry generators satisfy the $\mathfrak{u}(\mathcal{N})$ algebra
\be 
[\tensor{R}{^i_j},\tensor{R}{^k_\ell}]=\tensor{\delta}{^k_j} \tensor{R}{^i_\ell}-\tensor{\delta}{^i_\ell} \tensor{R}{^k_j}\ .
\ee

We now combine the super Poincar\'e algebra with the conformal algebra \eqref{eq:conformal algebra}. As in the bosonic case, we can compose inversion $\circ$ supercharge $\circ$ inversion to obtain a new fermionic generator. These new supercharges are conventionally denoted by $\tensor{S}{_i^\alpha}$ and $\bar{S}^{i\dot{\alpha}}$ and are called \emph{conformal supercharges}. Moreover, we are forced to include the R-symmetry generator $\tensor{R}{^i_j}$ in the algebra. This is very typical of SCFTs. In supersymmetric QFTs, R-symmetry acts by an outer automorphism on the algebra, but in SCFTs it is part of the superconformal algebra. Let us first write down the transformation behaviour of the supercharges under the bosonic subalgebra
\begin{subequations}
\begin{align} 
[\tensor{M}{_\alpha^\beta},\tensor{Q}{^i_\gamma}]&=\tensor{\delta}{_\gamma^\beta} \tensor{Q}{^i_\alpha}-\tfrac{1}{2} \tensor{\delta}{_\alpha^\beta} \tensor{Q}{^i_\gamma}\ , \\
[\tensor{M}{^{\dot{\alpha}}_{\dot{\beta}}},\tensor{\bar{Q}}{_{i\dot{\gamma}}}]&=\tensor{\delta}{^{\dot{\alpha}}_{\dot{\gamma}}} \tensor{\bar{Q}}{_{i\dot{\beta}}}-\tfrac{1}{2} \tensor{\delta}{^{\dot{\alpha}}_{\dot{\beta}}} \tensor{\bar{Q}}{_{i\dot{\gamma}}}\ , \\
[\tensor{M}{_\alpha^\beta},\tensor{S}{_i^\gamma}]&=-\tensor{\delta}{_\alpha^\gamma} \tensor{S}{_i^\beta}+\tfrac{1}{2} \tensor{\delta}{_\alpha^\beta} \tensor{S}{_i^\gamma}\ , \\
[\tensor{M}{^{\dot{\alpha}}_{\dot{\beta}}},\tensor{\bar{S}}{^{i\dot{\gamma}}}]&=-\tensor{\delta}{^{\dot{\gamma}}_{\dot{\beta}}} \tensor{\bar{S}}{^{i\dot{\alpha}}}+\tfrac{1}{2} \tensor{\delta}{^{\dot{\alpha}}_{\dot{\beta}}} \tensor{\bar{S}}{^{i\dot{\gamma}}}\ , \\
[D,\tensor{Q}{^i_\alpha}]&=\tfrac{1}{2} \tensor{Q}{^i_\alpha}\ , \\
[D,\tensor{\bar{Q}}{_i_{\dot{\alpha}}}]&=\tfrac{1}{2} \tensor{\bar{Q}}{_i_{\dot{\alpha}}}\ ,\\
[D,\tensor{S}{_i^\alpha}]&=-\tfrac{1}{2} \tensor{S}{_i^\alpha}\ , \\
[D,\tensor{\bar{S}}{^i_{\dot{\alpha}}}]&=-\tfrac{1}{2} \tensor{\bar{S}}{^i_{\dot{\alpha}}}\ , \\
[\tensor{R}{^i_j},\tensor{Q}{^k_\alpha}]&=\tensor{\delta}{_j^k} \tensor{Q}{^i_\alpha}-\tfrac{1}{4} \tensor{\delta}{^i_j} \tensor{Q}{^k_\alpha}\ , \\
[\tensor{R}{^i_j},\tensor{\bar{Q}}{_{k\dot{\alpha}}}]&=-\tensor{\delta}{_k^i} \tensor{\bar{Q}}{_{j\dot{\alpha}}}+\tfrac{1}{4} \tensor{\delta}{^i_j} \tensor{\bar{Q}}{_{k\dot{\alpha}}}\ , \\
[\tensor{R}{^i_j}, \tensor{S}{_k^\alpha}]&=-\tensor{\delta}{^i_k} \tensor{S}{_j^\alpha}+\tfrac{1}{4}\tensor{\delta}{^i_j} \tensor{S}{_k^\alpha}\ , \\
[\tensor{R}{^i_j}, \tensor{\bar{S}}{^k^{\dot{\alpha}}}]&=\tensor{\delta}{^k_j} \tensor{\bar{S}}{^i^{\dot{\alpha}}}-\tfrac{1}{4}\tensor{\delta}{^i_j} \tensor{\bar{S}}{^k^{\dot{\alpha}}}\ , \\
[\tensor{P}{_{\alpha\dot{\alpha}}},\tensor{S}{_i^\beta}]&=-2 \tensor{\delta}{_\alpha^\beta} \tensor{\bar{Q}}{_{i\dot{\alpha}}}\ , \\
[\tensor{P}{_{\alpha\dot{\alpha}}},\tensor{\bar{S}}{^{i\dot{\beta}}}]&=-2 \tensor{\delta}{_{\dot{\alpha}}^{\dot{\beta}}} \tensor{Q}{^i_\alpha}\ , \\
[\tensor{K}{^{\dot{\alpha}\alpha}},\tensor{Q}{^i_\beta}]&=2 \tensor{\delta}{_\beta^\alpha} \tensor{\bar{S}}{^{i\dot{\alpha}}}\ , \\
[\tensor{K}{^{\dot{\alpha}\alpha}},\tensor{\bar{Q}}{_{i\dot{\beta}}}]&=2 \tensor{\delta}{_{\dot{\beta}}^{\dot{\alpha}}} \tensor{S}{_i^\alpha}\ .
\end{align}
\end{subequations}
Hence supercharges have scaling dimension $+\tfrac{1}{2}$ and conformal supercharges have scaling dimension $-\tfrac{1}{2}$. This is compatible with the fact that (Poincar\'e) supercharges square to momentum generators. 

Finally, the new anti-commutators of the supercharges read
\begin{subequations}
\begin{align}
\{\tensor{Q}{^i_\alpha},\tensor{\bar{Q}}{_{j\dot{\alpha}}}\}&=\tfrac{1}{2} \tensor{\delta}{^i_j} P_{\alpha\dot{\alpha}}\ , \\
\{\tensor{\bar{S}}{^{i\dot{\alpha}}},\tensor{S}{_j^\alpha}\}&=\tfrac{1}{2} \tensor{\delta}{^i_j} K^{\dot{\alpha}\alpha}\ , \\
\{\tensor{Q}{^i_\alpha},\tensor{S}{_j^\beta}\}&=\tensor{\delta}{^i_j} \tensor{M}{_\alpha^\beta}+\tfrac{1}{2}\tensor{\delta}{^i_j} \tensor{\delta}{_\alpha^\beta}D-\tensor{\delta}{_\alpha^\beta} \tensor{R}{^i_j}\ , \\
\{\tensor{\bar{S}}{^i^{\dot{\alpha}}},\tensor{\bar{Q}}{_j_{\dot{\beta}}}\}&=\tensor{\delta}{^i_j} \tensor{\bar{M}}{^{\dot{\alpha}}_{\dot{\beta}}}+\tfrac{1}{2}\tensor{\delta}{^i_j} \tensor{\delta}{^{\dot{\alpha}}_{\dot{\beta}}}D+\tensor{\delta}{^{\dot{\alpha}}_{\dot{\beta}}} \tensor{R}{^i_j}\ .
\end{align}
\end{subequations}
While this looks complicated, the structure of the algebra is actually quite simple, since it defines a simple Lie superalgebra. We notice that the algebra can be understood as supermatrices of the form
\be 
\left(\begin{tabular}{c|c}
\text{conformal  algebra} & \text{supercharges} \\
\hline
\text{supercharges} & \text{R-symmetry}
\end{tabular}\right)
\ee
These matrices form a superalgebra called $\mathfrak{sl}(4|\mathcal{N})$. 

For $\mathcal{N}=4$, we notice that the generator $\tensor{R}{^i_i}$ is central\footnote{This means that it commutes with all elements of the algebra.} and can hence be consistently removed from the algebra. We call the resulting quotient $\mathfrak{psl}(4|4) \cong \mathfrak{sl}(4|4)/\mathfrak{u}(1)$ the $\mathcal{N}=4$ superconformal algebra. 

We summarise and reinstate the real form:
\begin{subequations}
\begin{align} 
\mathcal{N}\ \text{superconformal algebra} &\cong \mathfrak{su}(2,2|\mathcal{N})\ , \qquad \mathcal{N}=1,\, 2,\, 3\ , \\
\mathcal{N}=4\ \text{superconformal algebra} &\cong \mathfrak{psu}(2,2|4)\ .
\end{align}
\end{subequations}
Let us also summarise the representations of the bosonic subalgebra in which the supercharges transform:
\begin{subequations}
\begin{align}
\mathcal{N}&=1\ : & Q&\in [1;0]^{(-1)}_{\frac{1}{2}}\ , &  \bar{Q}&\in [0;1]^{(1)}_{\frac{1}{2}}\ , \\
\mathcal{N}&=2\ : & Q &\in [1;0]^{(1;-1)}_{\frac{1}{2}}\ ,& \bar{Q} &\in [0;1]^{(1;1)}_{\frac{1}{2}}\ , \\
\mathcal{N}&=3\ : & Q &\in [1;0]^{(1,0;-1)}_{\frac{1}{2}}\ ,& \bar{Q} &\in [0;1]^{(0,1;1)}_{\frac{1}{2}}\ , \\
\mathcal{N}&=4\ : & Q &\in [1;0]^{(1,0,0)}_{\frac{1}{2}}\ ,& \bar{Q} &\in [0;1]^{(0,0,1)}_{\frac{1}{2}}\ .
\end{align}
\end{subequations}
Here, we labelled representations by $[L]_\Delta^{(R)}$, where $L$ is the Lorentz representation label (specified by the Dynkin labels of the two $\mathfrak{su}(2)$'s), $\Delta$ is as before the scaling dimension and $(R)$ is the R-symmetry representation. It is determined by the $\mathcal{N}-1$ Dynkin labels of $\mathfrak{su}(\mathcal{N})$ and the overall $\mathfrak{u}(1)$ charge (except for $\mathcal{N}=4$). The last entry separated by a semicolon is the $\mathfrak{u}(1)$-charge.

Finally, we state the hermicity properties of the generators:
\begin{subequations}
\begin{align}
(\tensor{M}{_\alpha^\beta})^\dag&=\tensor{M}{_{\beta}^{\alpha}}\ , \\
(\tensor{M}{^{\dot{\alpha}}_{\dot{\beta}}})^\dag&=\tensor{M}{^{\dot{\beta}}_{\dot{\alpha}}}\ , \\
D^\dag&=D\ , \\
(P_{\alpha\dot{\alpha}})^\dag&=K^{\dot{\alpha}\alpha}\ , \\
(\tensor{R}{^i_j})^\dag&=\tensor{R}{^j_i}\ , \\
\big(\tensor{Q}{^i_\alpha}\big)^\dag&=\tensor{S}{_i^{\alpha}}\ , \\
\big(\tensor{\bar{Q}}{_i_{\dot{\alpha}}}\big)^\dag&=\tensor{\bar{S}}{^{i\dot{\alpha}}}\ ,
\end{align}
\end{subequations}
which can be motivated similarly as for the conformal algebra.
\question{\textbf{The hermitian conjugate.} Check that this definition of the dagger is consistent, in the sense that it is an anti-automorphism of the algebra:
\be 
[A^\dag,B^\dag]_{\pm}=[B,A]_{\pm}^\dag\ ,
\ee
where $[.,.]_{\pm}$ can be either a commutator or an anticommutator.}
\answer{Let us check some important (anti)commutators. We have for instance
\be 
[D^\dag,(\tensor{Q}{^i_\alpha})^\dag]=[D,\tensor{S}{_i^\alpha}]=-\tfrac{1}{2} \tensor{S}{_i^\alpha}\ . 
\ee
This has to be compared with
\be 
[D,\tensor{Q}{^i_\alpha}]^\dag=\tfrac{1}{2} (\tensor{Q}{^i_\alpha})^\dag=\tfrac{1}{2} \tensor{S}{_i^\alpha}\ ,
\ee 
which shows that this commutator respects the dagger.

Let us now look at the anti-commutators, which are the most important parts of the algebra. For the $\{Q,Q\}$ anticommutator, we have e.g.
\be 
\{(\tensor{Q}{^i_\alpha})^\dag,(\tensor{\bar{Q}}{_{j\dot{\alpha}}})^\dag\}=\{\tensor{S}{_i^\alpha},\tensor{\bar{S}}{^{j\dot{\alpha}}}\}=\tfrac{1}{2} \tensor{\delta}{^j_i} K^{\dot{\alpha}\alpha}
\ee
This has to be compared to
\be 
\{\tensor{Q}{^i_\alpha},\tensor{\bar{Q}}{_{j\dot{\alpha}}}\}^\dag=\tfrac{1}{2} (\tensor{\delta}{^i_j} P_{\alpha\dot{\alpha}})^\dag=\tfrac{1}{2} \tensor{\delta}{^j_i} K^{\dot{\alpha}\alpha}\ .
\ee
Finally, we look at the $\{Q,S\}$ anticommutator, for which we have
\be 
\{(\tensor{Q}{^i_\alpha})^\dag,(\tensor{S}{_j^\beta})^\dag\}=\{\tensor{S}{_i^\alpha},\tensor{Q}{^j_\beta}\}=\tensor{\delta}{^j_i} \tensor{M}{_\beta^\alpha}+\tfrac{1}{2}\tensor{\delta}{^j_i} \tensor{\delta}{_\beta^\alpha}D-\tensor{\delta}{_\beta^\alpha} \tensor{R}{^j_i}\ , 
\ee
which coincides with
\be 
\{\tensor{Q}{^i_\alpha},\tensor{S}{_j^\beta}\}^\dag=\left(\tensor{\delta}{^i_j} \tensor{M}{_\alpha^\beta}+\tfrac{1}{2}\tensor{\delta}{^i_j} \tensor{\delta}{_\alpha^\beta}D-\tensor{\delta}{_\alpha^\beta} \tensor{R}{^i_j}\right)^\dag\ .
\ee
Checking other (anti)commutators is very similar.
}

\subsection{\texorpdfstring{$d=6$}{d=6} superconformal symmetry}
Let us briefly turn to superconformal symmetry in six dimensions. We will again first consider the complexified algebra.

There is a similar `accident' for the conformal group as in four dimensions. The (complexified) conformal group is $\mathfrak{so}(8,\mathds{C})$. While there is no accidental isomorphism to another Lie algebra, $\mathfrak{so}(8,\mathds{C})$ enjoys triality symmetry, which is an outer automorphism that permutes the vector, the spinor and the cospinor representation. 

The putative supercharges should transform in a spinor representation of the conformal group, which up to a triality transformation, can also be viewed as the vector representation. There is one candidate superalgebra that satisfies this requirement, namely
\be 
\mathfrak{osp}(8|2\mathcal{N},\mathds{C})\ .
\ee
This is the orthosymplectic Lie algebra. It has the following structure:
\be 
\left(\begin{tabular}{c|c}
\text{orthogonal algebra} & \text{supercharges} \\
\hline
\text{supercharges} & \text{symplectic algebra}
\end{tabular}\right)\ ,
\ee
and the two off-diagonal blocks are identified. Thus, the R-symmetry is symplectic in this case. This is related to the fact that in six dimensions, the minimal spinors are symplectic Majorana-Weyl spinors. In particular, the two chiralities of spinors are independent of each other (whereas in four dimensions they are complex conjugates) and we can have $(\mathcal{N}_+,\mathcal{N}_-)$ supersymmetry. Conformal symmetry requires the supersymmetry to be of the form $(\mathcal{N},0)$. The maximal amount of supersymmetry is $\mathcal{N}=2$, since theories with higher $\mathcal{N}$ would not possess a stress-tensor multiplet.

The supercharges are in the representations
\begin{align} 
\mathcal{N}&=(1,0)\ : & Q &\in [1,0,0]_{\frac{1}{2}}^{(1)}\ ,  \\
\mathcal{N}&=(2,0)\ : & Q &\in [1,0,0]^{(1,0)}_{\frac{1}{2}}\ ,
\end{align}
of the bosonic subalgebra. Here the three Dynkin labels label the $\mathfrak{so}(6,\mathds{C}) \cong \mathfrak{sl}(4,\mathds{C})$ Lorentz representation. The superscripts are Dynkin labels of the R-symmetry algebra $\mathfrak{sp}(2\mathcal{N},\mathds{C})$. 

We can finally specify the real form. The real form of the bosonic subalgebra should be the triality image of $\mathfrak{so}(6,2) \oplus \mathfrak{sp}(2\mathcal{N})$. The corresponding real form is called $\mathfrak{so}(8^*)$ and hence the real form of the full superconformal algebra is called $\mathfrak{osp}(8^*|2\mathcal{N})$.
\subsection{Other dimensions} 
Let us briefly survey other dimensions. Here, we will only discuss the complexified algebras and not specify their real forms. It is understood that all appearing algebras are complex. Dimensions $d \ge 3$ are reviewed in \cite{Cordova:2016emh}. For two-dimensional superconformal algebras, see e.g.~\cite{Bershadsky:1986ms, Sevrin:1988ew}. It should now be clear what algebras qualify as  superconformal algebras. We are searching for simple Lie superalgebras with the following additional properties:
\begin{enumerate}
\item The bosonic subalgebra is $\mathfrak{so}(d+2) \oplus \text{R-symmetry}$. 
\item The supercharges transform in the spinor representation of the Lorentz subgroup $\mathfrak{so}(d)$.
\item The superalgebra should admit a suitable stress-energy tensor multiplet.
\end{enumerate}
We will discuss the stress-energy multiplet in Section~\ref{sec:4dSCFTs}.
In dimensions $d \ge 7$, there are no Lie superalgebras satisfying these constraints. This leads to the well-known fact that there are \emph{no} superconformal field theories in dimensions $d \ge 7$ \cite{Nahm:1977tg}. In $d=5$ there is one possibility, namely the exceptional Lie superalgebra $\mathfrak{f}_4$ with R-symmetry $\mathfrak{sl}(2,\mathds{C})$. It has half-maximal supersymmetry (namely eight $Q$-supercharges). 

In three dimensions, the relevant algebra is $\mathfrak{osp}(\mathcal{N}|4)$ and we can have $\mathcal{N}=1$, $2$, $3$, $4$, $5$, $6$ and $8$. 

In two dimensions, there is a plethora of possibilities. The conformal group factorises, $\mathfrak{so}(4)\cong \mathfrak{sl}(2,\mathds{C}) \oplus \mathfrak{sl}(2,\mathds{C})$, corresponding to left- and right-movers. Correspondingly, we can combine any left-moving superconformal algebra with any right-moving superconformal algebra. For the chiral superalgebras, there are the following choices.
 There are three infinite families $\mathfrak{sl}(2|\tfrac{1}{2}\mathcal{N})$ ($\mathcal{N}\in 2\mathds{Z}$ and $\mathcal{N}\ne 4$), $\mathfrak{osp}(\mathcal{N}|2)$ and $\mathfrak{osp}(3|\mathcal{N})$. Moreover, there are some exceptional possibilities. First, the case $\mathcal{N}=4$ is special in that there is actually a \emph{continuous family} of superconformal algebra given by the exceptional Lie superalgebra $\mathfrak{d}(2|1;\alpha)$. Finally, there is an exceptional $\mathcal{N}=7$ algebra given by $\mathfrak{g}_3$ and an exceptional $\mathcal{N}=8$ algebra given by $\mathfrak{f}_4$. The R-symmetries are $\mathfrak{g}_2$ and $\mathfrak{so}(7)$, respectively. We should note that in two dimensions, the full superconformal algebra is actually infinite-dimensional, since it is combined with Virasoro symmetry. What we have listed are the global subalgebras of these full superconformal algebras.
\section{Unitary representations}
Let us finally study unitary representations of these algebras, focussing on the case $d=4$. The main goal is to find the supersymmetric analogues of the unitarity bound \eqref{eq:unitarity bound 4d bosonic}. 
\subsection{Structure of superconformal representations}
We have seen that the conformal supercharges have scaling dimension $-\tfrac{1}{2}$ and the special conformal generators have scaling dimension $-1$. Because we again want to consider representations with bounded spectrum of $D$, we require that there is a state satisfying\footnote{Note that the anticommutator of the conformal supercharges yields the special conformal generator and thus the first condition is redundant.}
\be 
K^{\dot{\alpha}\alpha} \ket{[L]_\Delta^{(R)}}=0\ , \qquad \tensor{S}{_k^\alpha}\ket{[L]_\Delta^{(R)}}=0\ , \qquad \tensor{\bar{S}}{^k^{\dot{\alpha}}}\ket{[L]_\Delta^{(R)}}=0\ ,
\ee
where we wrote $(R)$ collectively for all R-symmetry labels. This state is called the \emph{superconformal primary}. Representations are then spanned by considering descendants obtained by acting with $Q$-supercharges and momenta. 

Because the supercharges are fermionic, one superconformal representation contains always finitely many conformal primaries. Thus, it is often useful to decompose the superconformal multiplet into conformal multiplets. The appearing conformal representations in long (i.e.~generic) supermultiplets follow from the representation theory of the bosonic subalgebra. For short representations, i.e.~representations saturating the unitarity bound, some conformal primaries might be absent.  

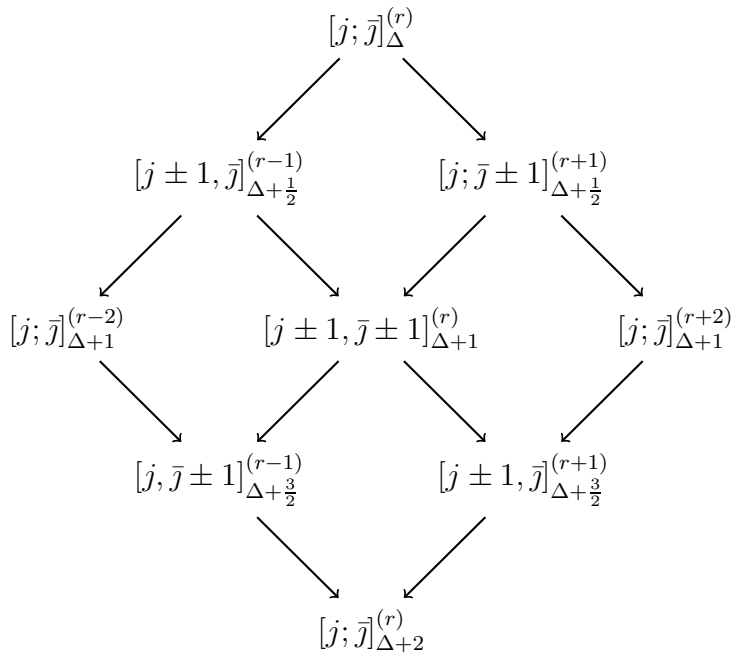
\begin{figure}
\begin{center}
\begin{tikzpicture}
\node (one) at (0,4) {$[j;\bar{\jmath}]_{\Delta}^{(r)}$};
\node (two left) at (-2,2) {$[j \pm 1,\bar{\jmath}]_{\Delta+\frac{1}{2}}^{(r-1)}$};
\node (two right) at (2,2) {$[j;\bar{\jmath}\pm 1]_{\Delta+\frac{1}{2}}^{(r+1)}$};
\node (three left) at (-4,0) {$[j;\bar{\jmath}]_{\Delta+1}^{(r-2)}$};
\node (three middle) at (0,0) {$[j\pm 1,\bar{\jmath}\pm 1]_{\Delta+1}^{(r)}$};
\node (three right) at (4,0) {$[j;\bar{\jmath}]_{\Delta+1}^{(r+2)}$};
\node (four left) at (-2,-2) {$[j ,\bar{\jmath}\pm 1]_{\Delta+\frac{3}{2}}^{(r-1)}$};
\node (four right) at (2,-2) {$[j\pm 1,\bar{\jmath}]_{\Delta+\frac{3}{2}}^{(r+1)}$};
\node (five) at (0,-4) {$[j;\bar{\jmath}]_{\Delta+2}^{(r)}$};
\draw[thick,->] (one) to (two left);
\draw[thick,->] (one) to (two right);
\draw[thick,->] (two left) to (three left);
\draw[thick,->] (two left) to (three middle);
\draw[thick,->] (two right) to (three right);
\draw[thick,->] (two right) to (three middle);
\draw[thick,->] (three middle) to (four right);
\draw[thick,->] (three right) to (four right);
\draw[thick,->] (three left) to (four left);
\draw[thick,->] (three middle) to (four left);
\draw[thick,->]  (four right) to (five);
\draw[thick,->]  (four left) to (five);
\end{tikzpicture}
\end{center}
\caption{The long $\mathcal{N}=1$ multiplet. The superconformal primary is the top state. The arrows show schematically the action of $Q$ and $\bar{Q}$ on the highest weight state. This is the generic situation for $j \ge 2$ and $\bar{\jmath} \ge 2$. For small values of $j$ and $\bar{\jmath}$, some of the conformal multiplets are missing according to the tensor product rules of $\mathfrak{su}(2)$.}\label{fig:N1 long multiplet}
\end{figure}

Let us exemplify this for 4d $\mathcal{N}=1$ superconformal symmetry. The descendants 
\be 
Q_\alpha \ket{[j;\bar{\jmath}]_\Delta^{(r)}}\qquad\text{and}\qquad \bar{Q}_{\dot{\alpha}} \ket{[j;\bar{\jmath}]_\Delta^{(r)}}
\ee
are again conformal primaries. They have bosonic representations $[j\pm 1,\bar{\jmath}]_{\Delta+\frac{1}{2}}^{(r-1)}$ and $[j;\bar{\jmath}\pm 1]_{\Delta+\frac{1}{2}}^{(r+1)}$. We can similarly continue and consider the descendants with two $Q$'s etc., which again lead to conformal primaries.\footnote{For two and more $Q$'s, the order of them matters and one might need to correct the state by a bosonic descendant to obtain a conformal primary. }
In this way, we generate $2^4=16$ conformal primaries within the superconformal multiplet. The structure of the multiplet is schematically displayed in Figure~\ref{fig:N1 long multiplet}.

For higher amounts of supersymmetry, there are in 4 dimensions in general $2^{4\mathcal{N}}$ states in a long multiplet. This number gets reduced for short multiplets saturating the unitarity bounds. 
\subsection{Unitarity bounds}
In general, the unitarity bounds have the following structure. Unitarity imposes that
\be 
\Delta \ge \Delta_A=f(j,\bar{\jmath},\text{R-symmetry})\ .
\ee
Representations saturating the unitarity bound have additional null vectors as in the bosonic case. In the supersymmetric case, the situation is richer than in the bosonic case. 
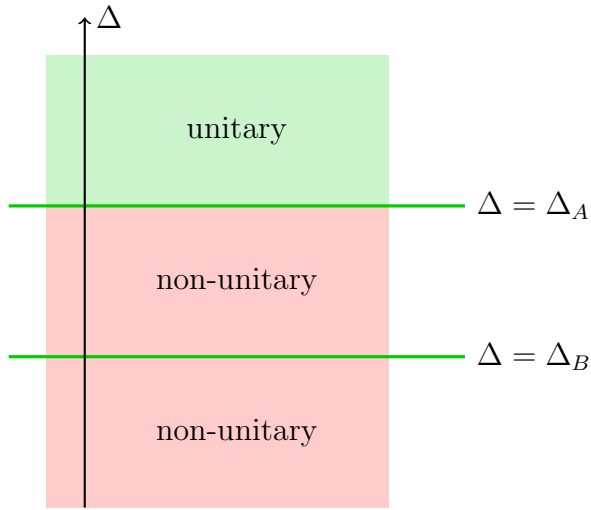
\begin{figure}
\begin{center}
\begin{tikzpicture}
\fill[red,opacity=.2] (-.5,-1) rectangle (4,3);
\fill[green!80!black,opacity=.2] (-.5,3) rectangle (4,5);
\draw[very thick,green!80!black] (-1,3) -- (5,3) node[right, black] {$\Delta=\Delta_A$};
\draw[very thick, green!80!black] (-1,1) -- (5,1) node[right,black] {$\Delta=\Delta_B$};
\draw[thick,->] (0,-1) -- (0,5.5) node[right] {$\Delta$};
\node at (2,2) {non-unitary};
\node at (2,0) {non-unitary};
\node at (2,4) {unitary};
\end{tikzpicture}
\end{center}
\caption{Unitarity structure of superconformal multiplets.}\label{fig:unitarity}
\end{figure}
Schematically, it is displayed in Figure~\ref{fig:unitarity}. Representations above the unitarity bound are in the green region. They do not have any null vectors and are long multiplets. Representations saturating the unitarity bound $\Delta=\Delta_A$ are still unitary, but have additional null vectors. Below the unitarity bound, there is a forbidden region. However, there can be further allowed short representations, which exist for discrete values of the scaling dimension. We will call this scaling dimension $\Delta_B$. The corresponding representations will be referred to as $L$ (for long), $A$ (if $\Delta_A$ is saturated) or $B$ (if the scaling dimension is $\Delta_B$). We have seen one very special instance of this already in the bosonic case, where the unitarity bound for a scalar particle in 4 dimensions reads $\Delta \ge 1$, but $\Delta=0$ also gives a unitary representation, namely the trivial representation.

This structure is very useful in practice. Often one only has access to a CFT at a particular point in its moduli space, for instance a weakly coupled Lagrangian description. We generally expect the scaling dimensions of the theory to change continuously as we change its parameters, i.e.~go for instance to strong coupling. Multiplets saturating the unitarity bounds are protected in the following sense. They can \emph{recombine} into long multiplets, whose scaling dimension is no longer protected. This is captured by the \emph{recombination rules}. We worked out one instance of a recombination rule for conformal symmetry in Exercise~\ref{recombination rules}. Analogue rules exist for superconformal symmetry, see e.g.~\cite{Cordova:2016emh} for a complete list. These protected multiplets modulo recombination are exactly captured by the superconformal index \cite{Romelsberger:2005eg, Kinney:2005ej}.
There are also \emph{absolutely protected} multiplets that can never recombine, in much the same way as the vacuum representation for conformal symmetry could never recombine. 

In four dimensions, there are two sets of supercharges, namely $Q$ and $\bar{Q}$ and similarly two spins $j$ and $\bar{\jmath}$. The superconformal multiplets can obey shortening conditions with respect to either the $Q$'s or the $\bar{Q}$'s or both. Because of this, the short multiplets can be actually of type $L\bar{L}$, $L\bar{A}$, $A\bar{L}$ etc.

\paragraph{$\mathcal{N}=1$ supersymmetry.}
Let us start with $\mathcal{N}=1$ supersymmetry. The $Q$-unitarity bounds are listed in Table~\ref{tab:Q bounds N1}. We also listed the conformal primary null vector, i.e.~the first conformal primary in the decomposition of the superconformal multiplet that becomes null.

\renewcommand{\arraystretch}{1.4}
\begin{table}
\begin{center}
\begin{tabular}{c|c|c|c}
name & primary & unitarity bound & conformal primary null state \\
\hline 
$L$ & $[j;\bar{\jmath}]_\Delta^{(r)}$ & $\Delta>2+j-\tfrac{3}{2}r$ & $-$ \\
$A_1$ & $[j;\bar{\jmath}]_\Delta^{(r)}$\ $(j \ge 1)$ & $\Delta=2+j-\tfrac{3}{2}r$ & $[j-1;\bar{\jmath}]_{\Delta+\frac{1}{2}}^{(r-1)}$ \\
$A_2$ & $[0,\bar{\jmath}]_\Delta^{(r)}$ & $\Delta=2-\tfrac{3}{2}r$ & $[0;\bar{\jmath}]_{\Delta+1}^{(r-2)}$ \\
$B_1$ & $[0,\bar{\jmath}]_\Delta^{(r)}$ & $\Delta=-\tfrac{3}{2}r$ & $[1,\bar{\jmath}]^{(r-1)}_{\Delta+\frac{1}{2}}$ 
\end{tabular}
\end{center}
\caption{The $Q$-unitarity bounds for four-dimensional $\mathcal{N}=1$ SCFTs.}\label{tab:Q bounds N1}
\end{table}
We then assemble the $Q$ and $\bar{Q}$-unitarity bounds to obtain all possibilities. This leads to Table~\ref{tab:representations N1}, which is the full list of possible unitary multiplets in 4d $\mathcal{N}=1$ SCFTs.
\renewcommand\tabcolsep{1pt}
\begin{table}
\begin{center}
\begin{tabular}{c|c|c|c|c}
& $\bar{L}$ & $\bar{A}_1$ & $\bar{A}_2$ & $\bar{B}_1$ \\
\hline
\multirow{2}{*}{$L$} & $[j;\bar{\jmath}]_\Delta^{(r)}$ & $[j;\bar{\jmath}\ge 1]_\Delta^{(r>\frac{j-\bar{\jmath}}{3})}$ & $[j;\bar{\jmath}]_\Delta^{(r>\frac{j}{3})}$ & $[j;\bar{\jmath}]_\Delta^{(r>\frac{j+2}{3})}$ \\
& $\scriptstyle{\Delta>2+\text{max}(j-\frac{3r}{2},\bar{\jmath}+\frac{3r}{2})}$ & $\Delta=2+\bar{\jmath}+\frac{3r}{2}$ & $\Delta=2+\frac{3r}{2}$ & $\Delta=\frac{3}{2}$\\ 
\hline 
\multirow{2}{*}{ $A_{1}$ } &$[j\geq 1;\bar{\jmath}]_{\Delta}^{(r<\frac{j-\bar{\jmath})}{3}}$& $[j\geq 1; \bar{\jmath}\geq 1]_{\Delta}^{(r=\frac{j-\bar{\jmath}}{3})}$ &  $[j\geq 1; \bar{\jmath}=0]_{\Delta}^{(r=\frac{j}{3})}$  &  $[j \geq 1; \bar{\jmath}=0]_{\Delta}^{(r=\frac{j+2}{3})}$  \\
& $\Delta=2+j-\frac{3r}{2}$ &$\Delta=2+\frac{j+\bar{\jmath}}{2}$ &$\Delta=2+\frac{j}{2}$ & $\Delta=1+\frac{j}{2}$ \\
\hline
\multirow{2}{*}{$A_{2}$} &$[j=0,\bar{\jmath}]_{\Delta}^{(r<-\frac{\bar{\jmath}}{3})}$& $[j=0; \bar{\jmath}\geq 1]_{\Delta}^{(r=-\frac{\bar{\jmath}}{3})}$ &$[j=0; \bar{\jmath}=0]_{\Delta}^{(r=0)}$ & $[j=0; \bar{\jmath}=0]_{\Delta}^{(r=\frac{2}{3})}$ \\
& $\Delta=2-\frac{3r}{2}$ &$\Delta=2+\frac{ \bar{\jmath}}{2}$&$\Delta=2$ &$\Delta=1$ \\
\hline
\multirow{2}{*}{$B_{1}$} &$[j=0;\bar{\jmath}]_{\Delta}^{(r<-\frac{\bar{\jmath}+2}{3})}$& $[j=0; \bar{\jmath} \geq 1]_{\Delta}^{(r=-\frac{\bar{\jmath}+2}{3})}$ & $[j=0; \bar{\jmath}=0]_{\Delta}^{(r=-\frac{2}{3})}$ & $[j=0;\bar{\jmath}=0]_{\Delta}^{(r=0)}$ \\
& $\Delta=-\frac{3r}{2}$ &$\Delta=1+\frac{ \bar{\jmath} }{2}$&$\Delta=1$ &$\Delta=0$ 
\end{tabular}
\end{center}
\caption{Multiplets for four-dimensional $\mathcal{N}=1$ SCFTs.}\label{tab:representations N1}
\end{table}
\renewcommand\tabcolsep{5pt}
\begin{table}
\begin{center}
\begin{tabular}{ c|c|c|c }
name &  primary &   unitarity bound & conformal primary null state  \\
\hline
$L$ & $[j;\bar{\jmath}]_{\Delta}^{(R ; r)}$  &$\Delta>2 + j+R-\frac{r}{2}$ & $-$ \\
$A_{1}$ & $[ j;\bar{\jmath}]_{\Delta}^{(R ; r)}$\  $(j\geq 1)$ &$\Delta= 2+j+R-\frac{r}{2}$ & $[j-1,\bar{\jmath}]_{\Delta+\frac{1}{2}}^{(R+1 ; r-1)}$ \\
$A_{2}$ & $[0,\bar{\jmath}]_{\Delta}^{(R;r)}$ &$\Delta=2+R-\frac{r}{2}$ & $[0,\bar{\jmath}]_{\Delta+1}^{(R+2; r-2)}$ \\
$B_{1}$ & $[0,\bar{\jmath}]_{\Delta}^{(R;r)}$& $\Delta=R-\frac{r}{2}$ & $[1,\bar{\jmath}]_{\Delta+\frac{1}{2}}^{(R+1; r-1)}$ 
\end{tabular}
\end{center}
  \caption{$Q$ shortening conditions in four-dimensional $\mathcal{N}=2$ SCFTs.} \label{tab:Q bounds N2}
\end{table} 

\paragraph{$\mathcal{N}=2$ supersymmetry.}
$\mathcal{N}=2$ supersymmetry behaves very similarly to $\mathcal{N}=1$ supersymmetry. The $R$-symmetry representation is now specified by the $\mathfrak{su}(2)$ Dynkin label $R$, as well as by the $\mathfrak{u}(1)$ charge $r$. The $Q$ shortening conditions are listed in Table~\ref{tab:Q bounds N2}. The $Q$- and $\bar{Q}$-shortening conditions can be combined to give the multiplets $X\bar{X}$, where $X\in \{L,\, A_1,\, A_2,\, B_1\}$ and similarly for $\bar{X}$. The full list of superconformal multiplets is given in Table~\ref{tab:representations N2}.

\question{\textbf{Shortening conditions for $\mathcal{N}=2$ supersymmetry.} \questionnumber{shortening conditions} The aim of this exercise is to derive parts of Table~\ref{tab:Q bounds N2}. There are three possible $Q$ shortening conditions of the $\mathcal{N}=2$ algebra. We analyse the shortening condition $B_1$
\begin{align}
\tensor{Q}{^1_\alpha} \ket{[0;\bar{\jmath}]_\Delta^{(R;r)}}&=0\ , \label{eq:first shortening condition}
\end{align}
where we also assumed $\ket{[0;\bar{\jmath}]_\Delta^{(R;r)}}$ to be a highest weight state under $R$-symmetry.
Derive the conditions on the scaling dimension $\Delta$ for this shortening to happen. Compare to Tables~\ref{tab:Q bounds N2} and \ref{tab:representations N2}.
}
\answer{
We have to compute the norm of this state. We have (no summation over $\alpha$)
\begin{align}
\left\lVert \tensor{Q}{^1_\alpha} \ket{[0;\bar{\jmath}]_\Delta^{(R;r)}} \right\lVert^2&= \bra{[0;\bar{\jmath}]_\Delta^{(R;r)}} \tensor{S}{_1^\alpha}  \tensor{Q}{^1_\alpha}  \ket{[0;\bar{\jmath}]_\Delta^{(R;r)}} \\
&=\bra{[0,\bar{\jmath}]_\Delta^{(R;r)}}  \tfrac{1}{2} D+\tensor{M}{_1^1}- \tensor{R}{^1_1} \ket{[0;\bar{\jmath}]_\Delta^{(R;r)}}  \ .
\end{align}
The three terms can be evaluated as follows. We have $D=\Delta$ when acting on the superconformal primary. By assumption, the superconformal primary does not transform under $\tensor{M}{_\alpha^\beta}$ and thus this term vanishes. Finally, we have to work out the correct normalisation of $\tensor{R}{^1_1}$. Consider the commutator
\be
[\tensor{R}{^1_1},\tensor{Q}{^k_\alpha}]=\begin{cases}
\tfrac{3}{4} \tensor{Q}{^k_\alpha}\ , \qquad &k=1\ , \\
-\tfrac{1}{4} \tensor{Q}{^k_\alpha}\ , \qquad &k=2\ .
\end{cases}
\ee
These values should be consistent with the fact that $Q$ has $r=-1$ and $R=1$. We thus see that $\tensor{R}{^1_1}=\tfrac{1}{2}R-\tfrac{1}{4}r$, where $R$ is the canonically normalised Cartan generator of $\mathfrak{su}(2)_R$. Thus, the norm becomes
\be 
\left\lVert \tensor{Q}{^1_\alpha} \ket{[0;\bar{\jmath}]_\Delta^{(R;r)}} \right\lVert^2=\tfrac{1}{2}\Delta-\tfrac{1}{2}R+\tfrac{1}{4} r\ .
\ee
Requiring positivity leads to the unitarity bound of Table~\ref{tab:Q bounds N2}.
}

\renewcommand{\arraystretch}{1.5}
\renewcommand\tabcolsep{2pt}
\afterpage{\clearpage}
\begin{sidewaystable}
\begin{center}
\begin{tabular}{ c|c|c|c|c}
 &  $L$ &  $A_1$ & $A_2$ & $B_1$  \\
\hline
\multirow{2}{*}{$L\ $} &$ [j;\bar{\jmath}]_{\Delta}^{(R;r)} $ & $ [j;\bar{\jmath}\geq 1]_{\Delta}^{(R;r>j-\bar{\jmath})}  $ &$ [j;\bar{\jmath}=0]_{\Delta}^{(R;r>j)} $&$[j;\bar{\jmath}=0]_{\Delta}^{(R;r>j+2)}$  \\
& $\Delta>2+R+\max\big\{j-\frac{1}{2}r,\bar{\jmath}+\frac{1}{2}r\big\}$ &$\Delta=2+R+\bar{\jmath}+\frac{1}{2}r$ &$\Delta=2+R+\frac{1}{2}r$ &$\Delta=R+\frac{1}{2}r$ \\
\hline
\multirow{2}{*}{$A_1\ $} &$ [j\geq 1;\bar{\jmath}]_{\Delta}^{(R;r< j-\bar{\jmath} )}  $& $ [j\geq 1;\bar{\jmath}\geq 1]_{\Delta}^{(R; r= j-\bar{\jmath} )} $ & $ [j\geq 1;\bar{\jmath}=0]_{\Delta}^{(R;r=j)} $ & $[j \geq 1;\bar{\jmath}=0]_{\Delta}^{(R;r=j+2)} $ \\
& $\Delta=2+R+j-\frac{1}{2}r$ &$\Delta=2+R+\frac{1}{2}(j+\bar{\jmath})$ &$\Delta=2+R+\frac{1}{2}\, j$ & $\Delta=1+R+\frac{1}{2}\, j$ \\
\hline
\multirow{2}{*}{$A_2\ $} &$ [j=0;\bar{\jmath}]_{\Delta}^{(R;r<- \bar{\jmath})} $& $ [j=0;\bar{\jmath}\geq 1]_{\Delta}^{(R;r=- \bar{\jmath})} $ &$ [j=0;\bar{\jmath}=0]_{\Delta}^{(R;r=0)} $&$ [j=0;\bar{\jmath}=0]_{\Delta}^{(R;r=2)} $ \\
& $\Delta=2+R-\frac{1}{2}r$ &$\Delta=2+R+\frac{1}{2}\, \bar{\jmath}$&$\Delta=2+R$ &$\Delta=1+R$ \\
\hline
\multirow{2}{*}{$B_1\ $} &$[j=0;\bar{\jmath}]_{\Delta}^{(R;r<-(\bar{\jmath}+2))}  $& $[j=0;\bar{\jmath} \geq 1]_{\Delta}^{(R;r=-(\bar{\jmath}+2))}  $ &$ [j=0;\bar{\jmath}=0]_{\Delta}^{(R;r=-2)} $&$ [j=0;\bar{\jmath}=0]_{\Delta}^{(R;r=0)} $ \\
& $\Delta=R-\frac{1}{2}r$ &$\Delta=1+R+\frac{1}{2}\, \bar{\jmath} $&$\Delta=1+R$ &$\Delta=R$ 
\end{tabular}
\end{center}
\caption{Multiplets for four-dimensional $\mathcal{N}=2$ SCFTs.}
\label{tab:representations N2}
 \end{sidewaystable} 

\paragraph{Other dimensions and supersymmetries.}
For other dimensions and supersymmetries, the situation is qualitatively similar and we refer to \cite{Cordova:2016emh} for details.

We will now discuss the differences appearing in six dimensions for both $\mathcal{N}=(1,0)$ and $\mathcal{N}=(2,0)$ supersymmetry. It is particular to four dimensions that we necessarily have both $Q$ and $\bar{Q}$ supercharges. This leads to two-sided shortening conditions discussed above. In six dimensions, there are only supercharges $Q$, because of the reality of the spinor. Thus, multiplets will be only labelled by one letter. On the other hand, there are more shortening conditions and besides $\Delta_B$, there are two more values $\Delta_D<\Delta_C<\Delta_B$ below the unitarity bound that lead to unitary multiplets. The corresponding multiplets are denoted by $L$, $A_i$ ($i=1,\dots,4$), $B_i$ ($i=1,\dots,3$), $C_i$ ($i=1,\, 2$) and $D_1$.

\section{4d \texorpdfstring{$\mathcal{N}=2$}{N=2} superconformal field theories} \label{sec:4dSCFTs}
In this section, we will exemplify the theory via important instances of multiplets in 4d $\mathcal{N}=2$ theories.
\subsection{Free multiplets}
\begin{figure}[h]
\begin{center}
\begin{tikzpicture}
\node (one) at (0,4) {$[0;0]_{1}^{(1;0)}$};
\node (two left) at (-2,2) {$[1;0]_{\frac{3}{2}}^{(0;-1)}$};
\node (two right) at (2,2) {$[0;1]_{\frac{3}{2}}^{(0;1)}$};
\draw[thick,->] (one) to (two left);
\draw[thick,->] (one) to (two right);
\end{tikzpicture}
\end{center}
\caption{The $\mathcal{N}=2$ (half)hypermultiplet.} \label{fig:4d N2 hyper}
\end{figure}
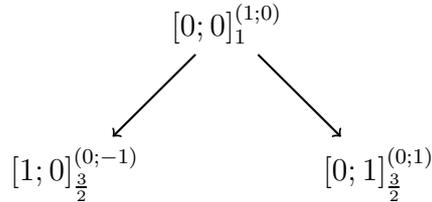

\paragraph{Hypermultiplet.} The basic matter multiplet of $\mathcal{N}=2$ supersymmetry is the hypermultiplet. It has the form of Figure~\ref{fig:4d N2 hyper}. We see that it is ultra-short and has $\Delta=1$. Thus, we see that in our classification it is the multiplet $B_1 \bar{B}_1[0;0]^{(1;0)}_1$. This multiplet is strictly-speaking the half-hypermultiplet. It does not appear alone in theories, since it is not CPT invariant and one has to add its CPT conjugate to get a full hypermultiplet.\footnote{This can be circumvented if the hypermultiplet is in a pseudo-real representation of the gauge and flavour groups. Since the R-symmetry representation $(1;0)$ is also pseudo-real, the scalars become real in this case.} As for the vector multiplet, the fields in the hypermultiplet satisfy the equations of motion of free fields.

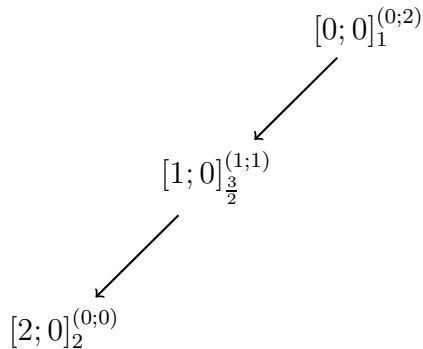
\begin{figure}
\begin{center}
\begin{tikzpicture}
\node (one) at (0,4) {$[0;0]_{1}^{(0;2)}$};
\node (two left) at (-2,2) {$[1;0]_{\frac{3}{2}}^{(1;1)}$};
\node (three left) at (-4,0) {$[2;0]_{2}^{(0;0)}$};
\draw[thick,->] (one) to (two left);
\draw[thick,->] (two left) to (three left);
\end{tikzpicture}
\end{center}
\caption{Half of the vectormultiplet.} \label{fig:4d N2 vector}
\end{figure}

\noindent
The full hypermultiplet is often written in terms of free fields as
\be 
\begin{tikzpicture}
\node (one) at (0,4) {$\psi^\alpha$};
\node (two left) at (-2,2) {$q$};
\node (two right) at (2,2) {$\tilde{q}^\dag$};
\node (three) at (0,0) {$(\tilde{\psi}^\alpha)^\dag$};
\draw[thick,->] (one) to (two left);
\draw[thick,->] (one) to (two right);
\draw[thick,->] (two left) to (three);
\draw[thick,->] (two right) to (three);
\end{tikzpicture}\ .
\ee
The fields of the hypermultiplet can be combined into two $\mathcal{N}=1$ chiral multiplets $q$, $\psi^\alpha$ and $\tilde{q}^\dag$, $(\tilde{\psi}^\alpha)^\dag$.
\paragraph{Vector multiplet.} The basic multiplet of 4d $\mathcal{N}=2$ gauge theories is the vector multiplet. The physical (i.e.~gauge invariant) fields are given by one anti-symmetric two-form $F_{\mu\nu}$, two complex Weyl fermions and one complex scalar. This multiplet is actually the sum of two irreducible superconformal representations. $F_{\mu\nu}$ can be further decomposed into its self-dual part $F^+_{\mu\nu}$ and its anti-self-dual part $F^-_{\mu\nu}$. The self-dual part transforms in the representation $[2;0]$ of the Lorentz group and the anti-self-dual part in the representation $[0;2]$. The resulting multiplet is displayed in Figure~\ref{fig:4d N2 vector}.\footnote{In a Lorentzian theory, the Hodge star on two-forms satisfies $\star_2^2=-1$. We declare $F_{\mu\nu}^\pm$ to satisfy $\star F^\pm_{\mu\nu}=\pm i F^\pm_{\mu\nu}$. Thus $F^\pm_{\mu\nu}$ are complex conjugates of each other. In a Euclidean theory, both $F^\pm_{\mu\nu}$ are independent and real.} Thus, reality of the action forces us to include both in the theory. 
In the above classification, they have the name $A_2\bar{B}_1[0;0]^{(0;2)}_{1}$ and $B_1\bar{A}_2[0;0]^{(0;2)}_{1}$. 

We should note that the appearing subrepresentations of the conformal group have themselves null-vectors. $F_{\mu\nu}$ has scaling dimension $\Delta=2$ in the above multiplet and hence has as a result of our classification of representations of the conformal group the null-descendant $\partial_\mu F^{\mu\nu}=0$. Similarly, the Weyl fermions $\lambda^\alpha$ and $\tilde{\lambda}^{\dot{\alpha}}$ satisfy the massless Dirac equation and the complex scalar $\phi$ obeys the massless Klein-Gordon equation $\partial_\mu \partial ^\mu \phi=0$.
This multiplet is hence often written as
\be 
\begin{tikzpicture}
\node (one) at (0,4) {$\phi$};
\node (two left) at (-2,2) {$\lambda^\alpha$};
\node (two right) at (2,2) {$\tilde{\lambda}^{\dot{\alpha}}$};
\node (three) at (0,0) {$F_{\mu\nu}$};
\draw[thick,->] (one) to (two left);
\draw[thick,->] (one) to (two right);
\draw[thick,->] (two left) to (three);
\draw[thick,->] (two right) to (three);
\end{tikzpicture}\ ,
\ee
which combines the two halves of the multiplet. These fields can be combined into an $\mathcal{N}=1$ vector superfield and an $\mathcal{N}=1$ chiral superfield.

\subsection{Conserved quantities}
\begin{figure}
\begin{center}
\begin{tikzpicture}
\node (one) at (0,4) {$[0;0]_{2}^{(2;0)}$};
\node (two left) at (-2,2) {$[1;0]_{\frac{5}{2}}^{(1;-1)}$};
\node (two right) at (2,2) {$[0;1]_{\frac{5}{2}}^{(1;1)}$};
\node (three left) at (-4,0) {$[0;0]_{3}^{(0;-2)}$};
\node (three middle) at (0,0) {$[1;1]_{3}^{(0;0)}$};
\node (three right) at (4,0) {$[0;0]_{3}^{(0;2)}$};
\draw[thick,->] (one) to (two left);
\draw[thick,->] (one) to (two right);
\draw[thick,->] (two left) to (three left);
\draw[thick,->] (two left) to (three middle);
\draw[thick,->] (two right) to (three right);
\draw[thick,->] (two right) to (three middle);
\end{tikzpicture}
\end{center}
\caption{The flavour current multiplet.} \label{fig:4d N2 flavour current}
\end{figure}
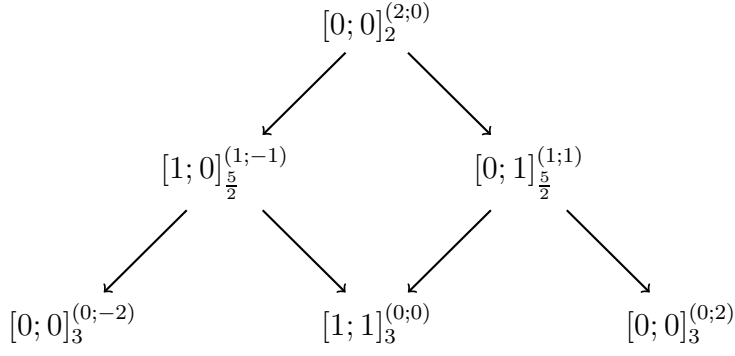

\paragraph{Flavour current multiplet.} Next, we discuss the multiplet of conserved flavour currents. Flavour currents $j^\mu$ give rise to flavour charges via
\be 
Q_\text{flavour}=\int_{\text{space}} \mathrm{d}^3 x\ j^0(x)\ .
\ee
By definition, flavour charges commute with the superconformal algebra (otherwise they would be R-symmetry generators). This means that we need
\be 
[Q,j^\mu]=\text{total derivative}\ .
\ee
Since total derivative means conformal descendant, we conclude that $j^\mu$ has to be the top component of the superconformal multiplet. This means that it is the (not necessarily unique) conformal primary within the multiplet with the largest scaling dimension. Moreover, $j^\mu$ is the vector representation $[1;1]$ and has to have scaling dimension $\Delta=3$, since $\partial_\mu j^\mu=0$ is a null descendant. Finally, $j^\mu$ is neutral under R-symmetry. The unique such multiplet is $B_1 \bar{B}_1[0;0]_2^{(2;0)}$ and is displayed in Figure~\ref{fig:4d N2 flavour current}. Often, the flavour symmetry carries some flavour symmetry index $a$ (if the flavour symmetry group is bigger than $\mathrm{U}(1)$). In this case, all components in the multiplet transform in the same flavour symmetry representation.
\paragraph{Stress tensor multiplet.} Finally, we discuss the stress tensor multiplet. We imposed its existence in the classification of superconformal algebras in Section~\ref{sec:superconformal symmetry}. Recall that the generators of the superconformal algebra arise as the integrals of local currents (we suppress R-symmetry indices for the moment):
\begin{subequations}
\begin{align}
P^\mu &= \int_\text{space} \mathrm{d}^3x\ T^{\mu 0}\ , \\
Q^{\alpha} &= \int_\text{space} \mathrm{d}^3x\ G^{0 \alpha}\ , \\
\bar{Q}^{\dot{\alpha}} &= \int_\text{space} \mathrm{d}^3x\ \bar{G}^{0 \dot{\alpha}}\ , \\
R^a &= \int_\text{space} \mathrm{d}^3 x\ j^{a,0}\ .
\end{align}
\end{subequations}
The other generators arise in fact from integrals of the \emph{same} local current, but with additional coordinate dependence:
\begin{subequations}
\begin{align}
M^{\mu\nu}&= \int_\text{space}\mathrm{d}^3x\ x^{[\mu} T^{\nu] 0}\ , \\
D&=\int_\text{space}\mathrm{d}^3 x\ x_\mu T^{\mu 0}\ , \\
K^\mu&=\int_\text{space}\mathrm{d}^3x\ (2x^\mu x^{\nu}-\eta^{\mu\nu} x^2) \tensor{T}{_\nu^0}\ , \\
\bar{S}^{\dot{\beta}}&= \int_\text{space} \mathrm{d}^3 x\ (\bar{\sigma}^\nu)^{\dot{\beta}\beta} x_\nu \tensor{G}{^0_\beta}\ ,\\
S^{\beta}&= \int_\text{space} \mathrm{d}^3 x\ (\sigma^\nu)^{\beta\dot{\beta}} x_\nu \tensor{\bar{G}}{^0_{\dot{\beta}}}\ .
\end{align}
\end{subequations}

These equations imply that the conserved (super)currents all sit in the same superconformal multiplet with $T_{\mu\nu}$ being the top component. This is the so-called stress tensor multiplet.

\begin{figure}
\begin{center}
\begin{tikzpicture}
\node (one) at (0,4) {$[0;0]_{2}^{(0;0)}$};
\node (two left) at (-2,2) {$[1;0]_{\frac{5}{2}}^{(1;-1)}$};
\node (two right) at (2,2) {$[0;1]_{\frac{5}{2}}^{(1;1)}$};
\node (three left) at (-4,0) {$[2;0]_{3}^{(0;-2)}$};
\node (three middle) at (0,0) {$[1;1]_{3}^{(2;0)}\oplus [1;1]_{3}^{(0;0)}$};
\node (three right) at (4,0) {$[0;2]_{3}^{(0;2)}$};
\node (four left) at (-2,-2) {$[1;2]_{\frac{7}{2}}^{(1;1)}$};
\node (four right) at (2,-2) {$[2;1]_{\frac{7}{2}}^{(1;-1)}$};
\node (five) at (0,-4) {$[2;2]_{4}^{(0;0)}$};
\draw[thick,->] (one) to (two left);
\draw[thick,->] (one) to (two right);
\draw[thick,->] (two left) to (three left);
\draw[thick,->] (two left) to (three middle);
\draw[thick,->] (two right) to (three right);
\draw[thick,->] (two right) to (three middle);
\draw[thick,->] (three middle) to (four right);
\draw[thick,->] (three right) to (four right);
\draw[thick,->] (three left) to (four left);
\draw[thick,->] (three middle) to (four left);
\draw[thick,->]  (four right) to (five);
\draw[thick,->]  (four left) to (five);
\end{tikzpicture}
\end{center}
\caption{The stress tensor multiplet in 4d $\mathcal{N}=2$ theories.} \label{fig:4d N2 stress tensor}
\end{figure}
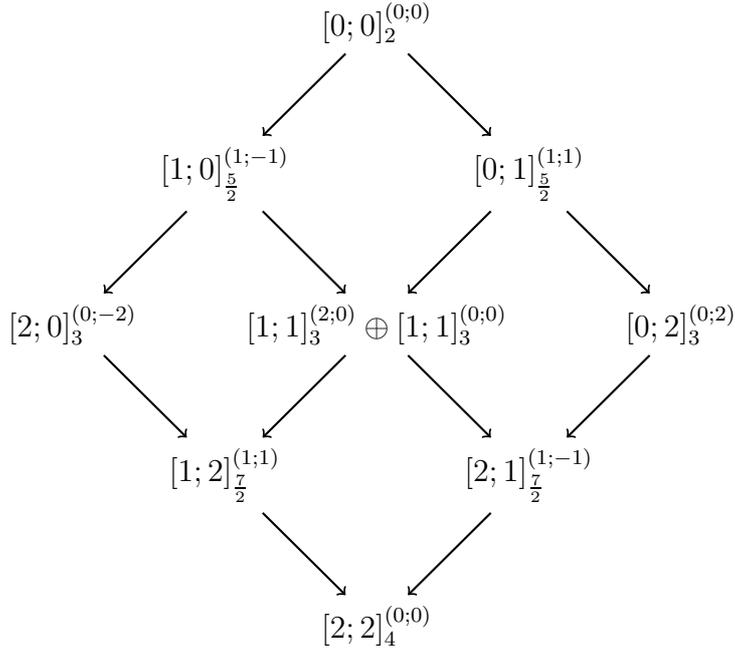

For $\mathcal{N}=2$ supersymmetry, the structure of the stress tensor multiplet is displayed in Figure~\ref{fig:4d N2 stress tensor}. The $\mathfrak{su}(2)\oplus \mathfrak{u}(1)$ R-symmetry currents are the central conformal primaries with $\Delta=3$. The supercurrents have $\Delta=\frac{7}{2}$ and the stress tensor is the top component with $\Delta=4$. This matches with the unitarity bounds from the conformal algebra and hence these are all automatically conserved. In the above classification, this multiplet is $A_2\bar{A}_2[0;0]^{(0;0)}_2$.
\subsection{Coulomb and Higgs branch operators}
 There is another important class of operators. For a Lagrangian $\mathcal{N}=2$ theory, one can analyse supersymmetric vacua. Let us denote the scalar of the vectormultiplet by $\phi$ and the two scalars of the hypermultiplet by $q_i$ and $\tilde{q}_i^\dag$ (here $i$ is a flavour index). These fields carry also gauge group indices, which we have suppressed. To obtain classical supersymmetric vacua, one has to minimize the scalar potentials of the theory, which results in the conditions
 \begin{subequations}
\begin{align}
0&=[\phi^\dagger,\phi]\ , \\
0&=\big(q_i q^{\dag,i}-\tilde{q}^\dag_i \tilde{q}^i\big)\Big|_{\text{traceless}}\ ,\label{eq:QQ condition} \\
0&=\phi^\dag q_i\ , \\
0&=\tilde{q}^i \phi^\dag\ .
\end{align}
\end{subequations}
There are two simple ways to satisfy these conditions:
\begin{enumerate}
\item $q=0$ and $\phi$ a normal matrix. This branch of solution is called the Coulomb branch. It breaks the gauge group $\text{G}$ generically down to $\mathrm{U}(1)^{\text{rank}(\mathrm{G})}$. We should note that $\phi$ is charged under $\mathfrak{u}(1)_r$, but uncharged under $\mathfrak{su}(2)_R$. 
\item $q\ne 0$ and $\phi=0$, subject to the condition \eqref{eq:QQ condition}. This branch of supersymmetric vacua is known as the Higgs branch.
\end{enumerate}
\paragraph{Coulomb branch operators and the chiral ring.} Operators that parametrise the Coulomb branch deformations are known as Coulomb branch operators. In a $\mathrm{SU}(n)$ gauge theory, these operators are
\begin{equation} 
\mathrm{tr}(\phi^k)\ , \qquad k=2,\dots,n\ . \label{eq:Phik monomials}
\end{equation}
In general, these operators preserve supersymmetry and hence sit in short representations. Moreover, they preserve $\mathfrak{su}(2)_R$ symmetry, but are charged under $\mathfrak{u}(1)_r$. Looking at Table~\ref{tab:representations N2}, we hence conclude that Coulomb branch operators are $L\bar{B}_1[0;0]^{(0;r)}$ for $r > 2$ and $A_2 \bar{B}_1[0;0]^{(0;2)}$ for $r=2$. They have $\Delta=\frac{r}{2}$ and are called \emph{chiral primaries}. Note that in principle, from the representation theory above, it is possible to have spinning Coulomb branch operators, i.e.~Coulomb branch operators that are not scalars. However, in all known theories, Coulomb branch operators are scalars and we shall assume this from now on.  There are also \emph{anti-chiral primaries} that transform in the representation $B_1\bar{L}[0;0]^{(0;r)}$ for $r > 2$ and $B_1\bar{A}_2[0;0]^{(0;r)}$  for $r=2$ and that satisfy $\Delta=-\frac{r}{2}$. These are the complex conjugate operators $\text{tr}(\bar{\phi}^k)$ (since $\phi$ is a complex scalar).

Coulomb branch operators are naturally endowed with a ring structure, the so-called \emph{chiral ring}. Their OPE has the form
\be 
\Phi_{r_1}(x_1) \Phi_{r_2}(x_2)=|x_1-x_2|^{\Delta-\Delta_1-\Delta_2} \Phi_{r_1+r_2}(x_2) +\dots\ ,
\ee
where $\Phi_{r_1}$ and $\Phi_{r_2}$ denote the chiral primaries and the dots denote terms that are more regular as $x_1 \to x_2$. By $\mathfrak{u}(1)_r$ conservation, $\Phi_{r_1+r_2}$ has charge $r_1+r_2$ and hence scaling dimension $\Delta\ge \frac{1}{2}(r_1+r_2)=\Delta_1+\Delta_2$. Thus, the exponent of $|x_1-x_2|$ cannot be negative. This means that one can define the ring structure by
\be 
(\Phi_{r_1}\cdot\Phi_{r_2})(x_2)\equiv \lim_{x_1\to x_2} \Phi_{r_1}(x_1) \Phi_{r_2}(x_2)\ ,
\ee
and $\Phi_{r_1} \cdot \Phi_{r_2}$ is again a chiral primary.
The chiral ring is typically freely generated, i.e.~does not have any algebraic relations. This is in particular true for $\mathrm{SU}(n)$ gauge theories, since the monomials \eqref{eq:Phik monomials} are algebraically independent.
\question{\textbf{Extremal correlators.} Consider a correlator of $n$ chiral primaries and one anti-chiral primary
\be 
\langle \Phi_{r_1}(x_1) \dots \Phi_{r_n}(x_n) \bar{\Phi}_r(x) \rangle\ ,
\ee
where $\sum_i r_i+r=0$. Such a correlator is called \emph{extremal correlator}. Show that its coordinate dependence is universally given by
\be 
\langle \Phi_{r_1}(x_1) \dots \Phi_{r_n}(x_n) \bar{\Phi}_r(x) \rangle= \text{const.} \times \prod_{i=1}^n |x_i-x|^{-2\Delta_i}\ .
\ee
}
\answer{ Let us first check that the proposed answer satisfies global Ward identities. It is clearly invariant under the Poincar\'e group. Thus, we only need to check the inversion, which is done as in Exercise~\ref{scalar three point function}.

Thus, we are free to use the conformal group to set $x \to \infty$. We define
\be 
\bar{\Phi}_r(\infty)\equiv \lim_{x\to \infty} |x|^{2\Delta} \bar{\Phi}_r(x)\ .
\ee
We then only have to prove that
\be 
\langle \Phi_{r_1}(x_1) \dots \Phi_{r_n}(x_n) \bar{\Phi}_r(\infty) \rangle
\ee
is coordinate independent. Take the derivative with respect to $x_i$. By the superconformal algebra, we can trade this derivative with $\{\tensor{Q}{^j_\alpha},\bar{Q}{_{k\dot{\alpha}}}\}$. We also know that
\be 
[\tensor{Q}{^j_\alpha},\bar{\Phi}_r(x)]=0\ , \qquad [\tensor{\bar{Q}}{^j_\alpha},\Phi_{r_i}(x_i)]=0\ .
\ee
For the anti-chiral primary, we have shown this in Exercise~\ref{shortening conditions}, the chiral primary is analogous. Since $R=0$ in this case, the proof of Exercise~\ref{shortening conditions}  actually works for both $\tensor{Q}{^j_\alpha}$, $j=1,2$. We can now use the (super)Jacobi identity to write
\be 
[\{\tensor{Q}{^j_\alpha},\bar{Q}{_{k\dot{\alpha}}}\},\Phi_{r_i}(x_i)]=\{\tensor{Q}{^j_\alpha},[\bar{Q}{_{k\dot{\alpha}}},\Phi_{r_i}(x_i)]\}+\{\bar{Q}{_{k\dot{\alpha}}},[\tensor{Q}{^j_\alpha},\Phi_{r_i}(x_i)]\}
\ee
As we have discussed the first term vanishes. Let us now insert this in the correlator
\begin{align}
&\langle \Phi_{r_1}(x_1) \dots[\{\tensor{Q}{^j_\alpha},\bar{Q}{_{k\dot{\alpha}}}\},\Phi_{r_i}(x_i)] \dots  \Phi_{r_n}(x_n) \bar{\Phi}_r(x) \rangle\nonumber\\
&\qquad=\langle \Phi_{r_1}(x_1) \dots\{\bar{Q}{_{k\dot{\alpha}}},[\tensor{Q}{^j_\alpha},\Phi_{r_i}(x_i)]\} \dots  \Phi_{r_n}(x_n) \bar{\Phi}_r(x) \rangle\\
&\qquad=\langle \Phi_{r_1}(x_1) \dots[\tensor{Q}{^j_\alpha},\Phi_{r_i}(x_i)]\dots  \Phi_{r_n}(x_n) [\bar{Q}{_{k\dot{\alpha}}},\bar{\Phi}_r(\infty)] \rangle\ ,
\end{align}
where in the last step, we have used the invariance of the correlator under supersymmetries (i.e.~the superconformal Ward identities) to integrate $\bar{Q}_{k\dot{\alpha}}$ by parts. Since it commutes with all the chiral primaries, it can only hit the anti-chiral primary. However, since the anti-chiral primary sits at $\infty$, this also vanishes: We have $[\bar{Q}{_{k\dot{\alpha}}},\bar{\Phi}_r(x)] \sim |x|^{-2(\Delta+\frac{1}{2})}$ inside the correlator for large $x$. But since  
\be 
\langle \Phi_{r_1}(x_1) \dots \Phi_{r_n}(x_n) \bar{\Phi}_r(\infty) \rangle=\lim_{x \to \infty} |x|^{2\Delta}\langle \Phi_{r_1}(x_1) \dots \Phi_{r_n}(x_n) \bar{\Phi}_r(x) \rangle\ ,
\ee
this does not survive in the limit. Hence
\be 
\partial_{x_i}\langle \Phi_{r_1}(x_1) \dots \Phi_{r_n}(x_n) \bar{\Phi}_r(\infty) \rangle=0\ ,
\ee
which proves the coordinate independence.
}
\paragraph{Higgs branch operators.} We can correspondingly define Higgs branch operators, which transform in short multiplets and are uncharged under $\mathfrak{u}(1)_r$, but sit in non-trivial $\mathfrak{su}(2)_R$ representations. Looking at Table~\ref{tab:representations N2}, these are the representations $B_1\bar{B}_1[0;0]^{(R;0)}$. By the same argument as above, they lead to a Higgs branch chiral ring. The Higgs branch operators satisfy typically many relations and correspondingly the Higgs branch chiral ring is \emph{not} freely generated.

\section*{Acknowledgements}
If is a pleasure to thank Elli Pomoni and Alessandro Sfondrini for organising the YIRSW school with such an interesting range of topics and for inviting me to give these lectures. I would also like to thank the other lecturers and all the participants of the school for many fruitful discussions. I acknowledge support from the Della Pietra Family at the IAS.

\bibliographystyle{JHEP}
\bibliography{bib}

\end{document}